\documentclass[pra,showpacs,twocolumn,floatfix, groupaddress,superscriptaddress,footinbib]{revtex4-1}
\usepackage{times,amsmath,amsfonts,amssymb,latexsym}
\usepackage{graphicx,epsf}
\usepackage{subfigure}
\usepackage{amsmath}
\usepackage{comment}

\usepackage{color}
\usepackage[normalem]{ulem}

\newcommand\R{{\mathrm {I\!R}}}

\newcommand{\be}{\begin{equation}}
\newcommand{\ee}{\end{equation}}
\newcommand{\ba}{\begin{eqnarray}}
\newcommand{\ea}{\end{eqnarray}}
\newcommand\tr{{\mbox{Tr\,}}}
\newcommand{\ignore}[1]{}

\newcommand{\ket}[1]{\left | {#1} \right \rangle }

\newcommand{\bra}[1]{\left \langle {#1} \right | }

\def\CC{{\rm\kern.24em \vrule width.04em height1.46ex depth-.07ex
    \kern-.30em C}}
\def\P{{\rm I\kern-.25em P}}
\def\RR{{\rm
         \vrule width.04em height1.58ex depth-.0ex
         \kern-.04em R}}
\def\bbbone{{\mathchoice {\rm 1\mskip-4mu l} {\rm 1\mskip-4mu l}
{\rm 1\mskip-4.5mu l} {\rm 1\mskip-5mu l}}}
\def\bbbc{{\mathchoice {\setbox0=\hbox{$\displaystyle\rm C$}\hbox{\hbox
to0pt{\kern0.4\wd0\vrule height0.9\ht0\hss}\box0}}
{\setbox0=\hbox{$\textstyle\rm C$}\hbox{\hbox
to0pt{\kern0.4\wd0\vrule height0.9\ht0\hss}\box0}}
{\setbox0=\hbox{$\scriptstyle\rm C$}\hbox{\hbox
to0pt{\kern0.4\wd0\vrule height0.9\ht0\hss}\box0}}
{\setbox0=\hbox{$\scriptscriptstyle\rm C$}\hbox{\hbox
to0pt{\kern0.4\wd0\vrule height0.9\ht0\hss}\box0}}}}
\def\bbbz{{\mathchoice {\hbox{$\sf\textstyle Z\kern-0.4em Z$}}
{\hbox{$\sf\textstyle Z\kern-0.4em Z$}}
{\hbox{$\sf\scriptstyle Z\kern-0.3em Z$}}
{\hbox{$\sf\scriptscriptstyle Z\kern-0.2em Z$}}}}

\usepackage{hyperref}

\begin{document}

\title{Random Quantum Batteries}
\author{Francesco Caravelli}
\affiliation{Theoretical Division and Center for Nonlinear Studies,\\
Los Alamos National Laboratory, Los Alamos, New Mexico 87545, USA}
\affiliation{Physics Department,  University of Massachusetts Boston,  02125, USA}
\author{Ghislaine Coulter-De Wit}
\affiliation{Physics Department,  University of Massachusetts Boston,  02125, USA}
\author{Luis~Pedro Garc\'{i}a-Pintos}
\affiliation{Physics Department,  University of Massachusetts Boston,  02125, USA}
\affiliation{Joint Center for Quantum Information and Computer Science,\
  NIST/University of Maryland,\
  College Park, Maryland 20742}
\affiliation{Joint Quantum Institute,\
  NIST/University of Maryland,\
  College Park, Maryland 20742}
\author{Alioscia Hamma}

\affiliation{Physics Department,  University of Massachusetts Boston,  02125, USA}
\affiliation{Univ. Grenoble Alpes, CNRS, LPMMC, 38000 Grenoble, France}

\begin{abstract}
Quantum nano-devices are fundamental systems in quantum thermodynamics that have been the subject of profound interest in recent years. Among these, quantum batteries play a very important role. In this paper we lay down a theory of random quantum batteries and provide a systematic way of computing the average work and work fluctuations in such devices by investigating their typical behavior. We show that the performance of random quantum batteries exhibits typicality and depends only on the spectral properties of the time evolving operator, the initial state and the measuring Hamiltonian. At given revival times a random quantum battery features a quantum advantage over classical random batteries. Our method is particularly apt to be used both for exactly solvable models like the Jaynes-Cummings model or in perturbation theory, e.g.,  systems subject to harmonic perturbations. We also study the setting of quantum adiabatic random batteries.
\end{abstract}

\maketitle

\section{ Introduction} Quantum batteries\cite{campaioli, alicki, pollini, PoliniPRB2019, PoliniPRB2018, PoliniPRL2019, Modiarxiv2017, LewensteinBatteries18} are a fundamental concept in quantum thermodynamics\cite{alicki0, aaberg2013truly, GelbwaserNJP2015, AlhambraPRX2016, 2019arXiv190202357A, levy, AdCPRL2017cycleengine, masanes2017general, CorreaPRE2013}, and they have attracted interest  as part of research in nano-devices that can operate at the quantum level \cite{anders2017focus, LindenPRL2010, workextractionPopescuNatComm14}. Tools and insights from quantum information theory have provided a natural bedrock for the description of quantum nano-devices and quantum batteries from the point of view of resource and information theory \cite{oppenheim,winter,LewensteinBatteries18, Huberreviewthermo2016, workextractionBrandaoPRL13,demkowicz2012NatComm,FrenzelPRE2014,HuberGaussianBatteries2017,gallego2016thermodynamic,infoscrambling}.

 In a closed quantum system, a battery can be modeled by a time-dependent Hamiltonian $H(t)$ evolving from an initial $H_0$ to a final $H_1$. The system is initialized in a state $\rho$ and, given that the entropy of the battery is constant under unitary evolution, the work extracted is given by the difference between the initial and final energies as measured in $H_0$ \cite{alicki}.

In this paper, we lay down the theory of Random Quantum Batteries (RQB). The randomness lies in  the initial state $\rho$, the Hamitonian defining the units of the energy $H_0$, and the time-evolution operator $U_t$. We are concerned with the average work extractable by (or storable in)  such a device and its fluctuations. 

The main results of this paper are: {\em (i)} proving a typicality result for the extracted work in a large class of time dependent quantum systems.  We show that - as the dimension $n$ of the Hilbert space becomes large - the extracted work is almost always given by the difference in energy between the initial state and the completely mixed state, amplified  by a {\em quantum efficiency} factor $1+Q_t/n^2$ that depends solely on the distribution of the eigenvalues of the exponential of the time-dependent perturbation operator $K$.  For $Q_t=0$, this result can be obtained by a classical system at infinite temperature. A random quantum battery can do it with limited energy resources. A non vanishing $Q$ is a contribution that is purely quantum and depends on the constructive interference between different eigenvalues of $K$.
The second main result is {\em (ii)} to provide a general method to study the average extractable work and its fluctuations in perturbation theory, which is essential to obtain results for physically relevant systems beside  few exactly solvable models. We study as an example the Jaynes-Cummings model with a harmonic perturbation.  
Finally {\em (iii)}, we study the case of adiabatic random quantum batteries, that is, batteries that operate slowly, so that there is no inversion of the populations of the energy levels. We show that also  adiabatic random quantum batteries feature typicality in the large Hilbert space dimension $n$ limit. 

There is a large interest in typical properties in batteries due to the effect of disorder and the environment. In \cite{Ghosh} a model of quantum battery based on a spin chain is studied where randomness is introduced as disorder in the couplings of the Hamiltonian $H_0$. In \cite{Andolina} the disorder is introduced in the interaction Hamiltonian  which is chosen to be in the Many-Body Localized (MBL) phase.
In \cite{adc,adc2}, the work statistics in the scenario of a random quantum quench are computed, and it is shown that the knowledge of the work statistics in this setting yields information on the Loschmidt echo dynamics. The importance of work fluctuations in quantum thermodynamics in a different setting than ours was also studied in \cite{sss}.

 %%%%%%%%%%%%%%%%%%%%%%%%%%%%%%%%%%%%%%%%%%%%%%%%%%%%%%

%%%%%%%%%%%%%%%%%%%%%%%%%%%%%%%%%%%%%%%%%%%%%%%%%%%%
% FIGURE
%%%%%%%%%%%%%%%%%%%%%%%%%%%
\ignore{
\begin{figure}
  \centering
  \includegraphics[scale=.4]{figure.pdf}
  \caption{}
  \label{fig1}
\end{figure} 
}
%%%%%%%%%%%%%%%%%%%%%%%%%%%

%%%%%%%%%%%%%%%%%%%%%%%%%%%%%%%%%%%%%%%%%%%%%%%%%%%%%%

%%%%%%%%%%%%%%%%%%%%%%%%%%%%%%%%%%%%%%%%%%%%%%%%%
%SECTION
%%%%%%%%%%%%%%%%%%%%%%%%%%%%%%%%%%%%%%%%%%%%%%%%%

\section{Setup} 
In this section, we are studying the typical behavior of random batteries when the interaction Hamiltonian  is a random operator. The importance of this approach lies in the fact that typicality is a powerful argument to establish general features in quantum thermodynamics. As an example, typicality of entanglement in Hilbert space can be used to explain thermalization in a closed quantum system\cite{popescu}. On the other hand, this approach is useful to argue about the robustness of a model of  quantum battery.

We model the quantum battery in the following way. Start with a finite dimensional Hilbert space $\mathcal H = \mathbb{C}^n$, and time-dependent Hamiltonians $H(t)\in \mathcal{B(H)}$, that is, a bounded Hermitian operator on $\mathcal H$. The initial state of the system will be denoted by $\rho$ and its time evolution by $\rho_t = \mathcal U_t \rho \equiv U_t\rho U^\dagger_t$, where the unitary evolution operator is given by the time-ordered product $U_t = \mathcal T \exp(-i\int^t_0 H(s)ds)$. We  model the Hamiltonian in two  ways. In the first scenario we consider the time-dependence as a perturbation of a time-independent Hamiltonian $H_0$, that is, $H_G(t) = H_0 + V_G (t)$. The subscript $G$ indicates the randomness of the perturbation which we take to be $V_G (t)= G^\dagger V(t) G$, where $G$ is a unitary representation of the unitary group on $\mathbb{C}^n$. In the second scenario we consider the time evolution generated by adiabatic evolution induced by a Hamiltonian $H_G(t)$, where the $G_t$ is a family of unitary operators that rotates the projectors onto the subspaces of a given energy. The discussion of the adiabatic scenario is deferred to section \ref{adiabscen}.

In both settings, we can similarly model randomness in the initial state $\rho$ or Hamiltonian $H_0$ also by random rotations $\rho_G = G\rho G^\dagger$ and $H_G =G^\dagger H_0 G$. 
Loosely speaking, we will refer to the spectra of the initial state, of the measuring Hamiltonian $H_0$, and of the evolution operator 
\ba
K = \mathcal T \exp \left(-i\int^t_0 V(s) ds\right)
\ea
collectively as the {\em battery spectrum}. 
Notice that all these randomizations preserve the battery spectrum. This is a crucial point in this paper, as we are interested in ensembles of quantum batteries with a given spectrum. Randomizing also over the spectrum will yield, as we shall see, trivial results. 

In our setting the system is closed and evolves unitarily and the entropy of the battery does not change. Thus the work extracted from the quantum battery is given by 
\ba
W(t) =\tr[(\rho-\rho_t)H_0]
\ea
(or \textit{ergotropy} \cite{campaioli,erg}).

As mentioned before, this approach is different from the type of disorder in the couplings considered in the literature\cite{Ghosh,Andolina,adc,adc2}, as for us disorder is a random rotation $G$ that mantains the spectrum of the eigenvalues of the perturbation $V_G (t)$ (the interaction). 

{
A simple example which clarifies how our disorder affects the extracted work $W$ is the following single spin case inspired by  nuclear magnetic resonance (NMR). We consider a Hamiltonian of the form $H_0= \hat \sigma_x+H_{int}$, where $H_{int}=g \vec \sigma\cdot \vec B (t)$, and $\vec B(t)=(b_x(t),b_y(t),b_z(t))$ is the external magnetic field and $\vec \sigma$ the Pauli matrices. The spectrum of the interaction is effectively dependent only on the norm of the external field $\vec B$, which can  however be directed in all directions. We focus on an average which keeps the spectrum of the interaction constant, but rotates its basis. A two-level system example is the Jaynes-Cummings model of optics, on which we focus our attention in a random electromagnetic background. Precise experiments in these systems exist and thus provide a good background for testing the typical behavior of (random) quantum batteries \cite{Nat1,Nat2}.}

 The Hamiltonian $H_0$  defines the energy measurement, that is, the amount of energy stored in the battery.
If we had access to any possible random Hamiltonian $H(t)$, we would expect that the average state $\rho_t$ after the evolution should be the completely mixed state, in which case the average work extracted would be $\langle{W}\rangle = E_0 -\tr H_0/n$. 
This work is positive (that is, the battery has discharged) if the initial energy was larger than the energy in the completely mixed state, or it has charged if the initial state was populating the lower levels of $H_0$. Notice that this setting we have arbitrary hamiltonians $H(t)$ that can access arbitrary high energies as measured by $H_0$. Instead, we  ask how much  work can be extracted  if we have limited energetic resources, that is, when the spectra of $H_0$ and $V(t)$ are fixed. This motivates our setting in terms of rotations of the time dependent part of the Hamiltonians as $H_G(t) = H_0 + V_G (t)$. 

In the following, we are interested in the average work obtained by averaging over initial states $\rho$,  the measurement of energy  Hamiltonian $H_0$, and the time dependent Hamiltonian $H_G(t)$. The averages are performed according to the Haar measure on $\mathbb{C}^n$. The fluctuations of work are defined through the same Haar averaging as $\Delta W^2 =\langle (W-\langle W\rangle)^2\rangle$. In the following, the symbol $\langle X\rangle$ will represent the Haar average $\langle X\rangle= \int dU G_U^\dagger  X G_U$, where $G_U$ is the suitable representation of the unitary group.  We use standard techniques for the Haar averaging (see e.g. \cite{lis, hammaavg,hammaavg2}) to compute the average and variances according to the Haar measure.

\subsection{Work and quantumness} 
A quick calculation shows that $W(t) = \tr \{ U^\dagger_t H_0 [\rho, U_t]\} = \tr\{\rho [U_t, U^\dagger_t H_0]\}=\tr \{U_t[U^\dagger_tH_0, \rho]\}$. These expressions imply that the extractable work depends on the lack of commutativity between the initial state $\rho$, the evolution operator $U_t$, and the Hamiltonian $H_0$. Moreover, they show that the coherence of the initial state in the eigenbasis of the evolution operator is necessary to have non vanishing extractable work from a quantum battery\cite{horod, FrancicaPRE2019, workextractionAbergPRL14, brandner2015coherence, korzekwa2016extraction, PetruccioneSciRep2019, lostaglio2015quantum, MarvianPRA2016, CoherenceRevMod2017}. In particular, if the initial state is a steady state for the unitary evolution, the work is identically zero and so are work fluctuations. It is interesting that coherence in two different bases plays a role, which calls for a multi-basis definition of coherence from
 the resource theoretic point of view.  In the following, we will see that this lack of commutativity takes the form of out of time order correlators, which is a hint to the connection between performance of quantum batteries and quantum chaos~\cite{adc}. 
 Notice that these expressions are also valid in the interaction picture $U_I = \exp (iH_0 t) U_t = \mathcal T \exp (-i \int^t_0 V_G(s) ds )  = G \mathcal T \exp (-i\int^t_0 V(s) ds) G^\dagger\equiv GKG^\dagger$, an expression that will become useful later. Bounds on the stored and extracted energy have been obtained recently in \cite{Riera}, also in terms of the quantum Fisher information for the power $P_t=\frac{d}{dt} W$.

As we remarked above, with no limit on energetic resources one can bring the system on average in the completely mixed state. A quantum channel that just dephases the system and mixes up the populations can achieve the same final result. The same result can be obtained by a classical system working at infinite temperature. Consequently,  we are also interested in whether quantum coherence plays a specific role in outperforming the mixed state case. As we shall see, partial revivals due to the build-up of quantum coherence provide a quantum advantage.

\section{Average work and fluctuations in RQBs} 
In this section, we show how the average work and its fluctuations behave in quantum random batteries when we randomize over the initial states $\rho$, the measuring Hamiltonian $H_0$, or the interaction $V(t)$. In all cases, this average is obtained by rotating these operators by a random unitary operator and by taking the Haar average.

Let us start by computing the average work obtained by a generic quantum evolution and averaging over all the initial states. It should not be surprising that the average extracted work amounts to zero. Indeed, we have
\ba\nonumber
\langle W\rangle_\rho &=&\tr \left[ \langle \rho\rangle (H_0 - U^\dagger H_0 U) \right] \\
&=&\frac{1}{n}  \bbbone \tr \delta H_0  =0
\ea
where we defined the traceless operator $\delta H_0\equiv \left[  H_0 - U^\dagger H_0 U \right]$ and have used  that the Haar-average state in the Hilbert space is $ \langle \rho\rangle = 1/n\ \bbbone$. 
However, the fluctuations are not trivial\cite{MasanesNatComm2016work}. Details of the calculation are in \ref{appendix1}. We obtain
\ba
%\nonumber \Delta W^2_\rho &=& \frac{n\tr\rho^2-1}{n(n^2-1)}\tr \delta H_0^2 \\
\Delta W^2_\rho&=& 2\frac{n\tr\rho^2-1}{n(n^2-1)}(\tr H_0^2-\tr(H_0U_t^\dagger H_0 U_t))
\ea
 It is remarkable that the maximum of the fluctuations are reached for a pure state whereas they decrease with the purity of the initial state, and are identically zero if the system is initialized in the completely mixed state. Similarly, fluctuations in the work are smaller the larger the fluctuations in the eigenvalues of $H_0$. Notice that the time dependent part has the form of a (two-point) {out-of-time-ordered correlator (OTOC)\cite{otocs, yoshida}.
 
What happens instead if we choose randomly the measuring Hamiltonian $H_0$? As we said above, we model this family of Hamiltonians as $H_G =G^\dagger H_0 G$. This is a sensible definition as it gives us results that still  depend on the spectrum of the Hamiltonian. Again, it should not surprise that the average work is zero, since 
\ba
\langle W\rangle_{H_0} = \tr  \left[ ( \rho-\rho_t) \langle H_0 \rangle\right] = \frac{\tr H_0}{n} \tr(\rho-\rho_t)=0
\ea
as the average of every operator in the trivial representation is proportional to the identity, and $\rho-\rho_t$ is traceless. Some tedious calculations in Appendix \ref{appendix2} show that the work fluctuations are given by 
\ba
\Delta W^2_{H_0}  =  \langle W^2\rangle_{H_0} =  \frac{2n}{n^2-1 }\Delta H_0^2\tr(\rho^2-\rho\rho_t)
\ea
where $\Delta H_0^2 = \frac{1}{n} \tr H^2_0 -\frac{1}{n^2}(\tr H_0)^2$ are the fluctuations of the eigenvalues of $H_0$, namely the fluctuations of $H_0$ in the completely mixed state. Again, the time-dependent part $\tr(\rho\rho_t)$ has the form of an OTOC. The connection between OTO correlators and Loschmidt echo has recently been investigated in \cite{infoscrambling}. 
In terms of the $2-$norm fidelity $\mathcal F_2 (\rho,\sigma)= \tr(\rho\sigma)/\max [\tr\rho^2,\tr\sigma^2]$ and the Loschmidt echo $\mathcal L_t = \mathcal F_2 (\rho\rho_t)$, we have 
\ba
\Delta W^2_{H_0}  &=& \frac{2n}{n^2-1 }\Delta H_0^2\tr\rho^2[1-\mathcal L_t] 
\ea
Notice that as $\mathcal L_t$ is typically scaling as $n^{-2}$\cite{zanardicampos}, the average fluctuations are determined only by the fluctuations in $H_0$ and the purity of the initial state. However, at specific, revival times, there is a spike in fluctuations. 
Moreover, if we consider the average work over a large time $T$, the average Loschmidt echo becomes the purity of the the completely dephased state in the basis of the Hamiltonian, $\bar{\rho}$  and the above expression reads
\ba
\overline{\Delta W^2}^T  = \frac{2n}{n^2-1 }\Delta H_0^2\tr\rho^2(1-\tr\bar{\rho}^2)
\ea
where the time average over a time $T$ is defined as $\overline{f}^T\equiv T^{-1}\int_0^T  f(t) dt $. 
 We see that large fluctuations can be achieved if there are not only large fluctuations in the energy gaps of the Hamiltonian $H_0$, but also if the initial state is pure enough, or if the time evolution is nontrivial.
  If  the initial state is very mixed or the time evolution does not feature an exponentially decaying Loschmidt echo, then work fluctuations will be negligible regardless of $H_0$. %The average work over long time is maximized by choosing a pure state that is very spread out in the eigenstates of $H_0$. 

At this point, we are ready to tackle our main goal, that is, to compute the work and its fluctuations in a quantum battery modeled by $H_G(t) = H_0 + V_G (t)$. In this setup, one has perfect control on the measuring Hamiltonian, but the controlled quantum evolution is very noisy, as $V_G (t)= G^\dagger V(t) G$. However, one has retained control on the spectrum of the driving Hamiltonian, which is an experimentally realistic situation. 
In the interaction picture, and by defining $C\equiv \tr [U_I\rho U_I^\dagger H_0]$, we see that work is given by 
\ba\nonumber
W(t) &=& %\tr[(\rho-\rho_t) H_0] = 
\tr[\rho H_0] - \tr[\rho_t H_0] \equiv E_0 - \tr[\rho_t H_0] \\
&=& E_0 - \tr [U_I\rho U_I^\dagger H_0]\equiv E_0- C
\ea
We can write the above expression as
\ba
W(t) &=& E_0 -\tr\left[  U_I \rho\otimes U_I^\dagger H_0 T^{(2)}  \right] \\
&=& E_0- \tr\left[(\rho \otimes H_0) (U_I\otimes U_I^\dagger)  T^{(2)} \right]
\ea
Now recall that the interaction picture operator $U_I$ depends on the random rotations $G$ as $GKG^\dagger$. The average work $\langle W(t)\rangle_V$ over the noise $G$ can then be computed (see Appendix \ref{appendix1} for details) to give 
\ba\label{avwork}
\langle W(t)\rangle_V = \left[ E_0-\frac{\tr H_0}{n}\right] \left[ \frac{n}{n+1} +\frac{Q_t}{n^2-1}\right]
\ea
with
\begin{eqnarray}\label{WV2}
    Q_t &=& -2\sum_{j\neq k} \cos(\theta_j-\theta_k)
\end{eqnarray}
where $\lambda_k = \exp(i\theta_k)$ are the eigenvalues of the evolution operator $K=\mathcal T \exp (-i\int^t_0 V(s) ds) $.  The time dependence of the work is thus contained in the function $Q_t$. 
For large dimension $n$, the average work reads 
\begin{eqnarray}
    \langle W(t)\rangle_V &=&\left(E_0-\frac{\tr H_0}{n}\right)\left(1+\frac{Q_t}{n^2}\right) \nonumber \\
    &=& \tr[(\rho-\bbbone/n)H_0)](1+Q_t/n^2).
\end{eqnarray} 
At this point, averaging over the initial state $\rho$ would give zero, while averaging over the Hamiltonian $H_0$ gives an exponentially small work $\sim n^{-1}$. 

 Let us comment on the meaning of the result Eq.(\ref{avwork}). We are starting with an initial state $\rho$ and evolving with a random evolution generated by $V(t)$. So far we have averaged over rotations of the time dependent perturbation $V_G(t)$. Such rotations keep the eigenvalues of $V_G$ unchanged so that all the results are a function of spectral quantities like $Q_t$.  One could expect that, if the evolution were completely random, one would end up with the completely mixed state, and then the work extracted would have to be $W= (E_0-\tr H_0/n)$. However,  we have fixed the spectrum of $V(t)$ in the randomization, so it is remarkable that one can achieve the infinite temperature result.

 Moreover, in the average work $\langle W(t)\rangle_V$ there is an amplifying  quantum correction $(1+Q_t/n^2)$. 
These corrections are quantum in nature because they correspond to the constructive interference that builds up in $Q_t=- 2\sum_{j\ne k}\cos(\theta_j-\theta_k)$. One expects that without a specific structure in the $\theta$'s, the factor $Q_t/n^2$ would rapidly decay to zero. This means that on average (and typically) one can achieve in this setting the same result that would be attained with random arbitrary resources. However, we can do better than that. First, if fluctuations are not a concern,  
it is possible for nano-systems with small $n$ to have large $Q_t$.
We are going to give an example in the following, using an optical cavity. Moreover, it is possible to design devices with a spectrum such that,
for specific values of $t$, the term $Q_t$ is of order one, which can be exploited as quantum advantage in the construction of a battery. In the next section, we show how, in a specific example,
revivals in $Q_t$ allow the battery to outperform
the infinite temperature (and classical) behavior. 

The question of what happens in the large $n$ case is very interesting. In the optical cavity application shown in the next section IV.A, the quantum amplifying factor is washed out as $n^{-2}$. We think that this would happen for most models. In this sense, this is a sign of the loss of quantumness as the dimension of the Hilbert space grows. One wonders, though, whether for some specific model the amplifying factor $Q_t/n^2$ might not disappear in the large $n$ limit. Finding such a realistic model would be of enormous practical interest. Conversely, proving that no model can feature this advantage as $n$ goes to infinity would be a very interesting result in quantum thermodynamics. 

As mentioned, one expects that for a random matrix its spectrum should yield a vanishing $Q_t$. 
A natural question to ask then is what the
typical behavior of this quantity is
when these eigenvalues are taken randomly, according to a CUE distribution, 
(see e.g. \cite{adc}). 
Let us define $r_k={\lambda_{k+1}}/{\lambda_k}$. 
We prove in Appendix \ref{appendix8e} that
\begin{eqnarray}
    Q&=&\frac{1}{2} \sum_{k=1}^n \sum_{j=k+1}^n \left( \prod_{i=j+1}^n r_i +\prod_{i=j+1}^n r_i^{-1}\right).
\end{eqnarray}
The behavior of $Q$, evaluated numerically, is depicted in Fig. \ref{fig:cue}. We see that for large $n$ the peak of the distribution moves towards zero. That is, averaging over the spectra does not give any amplification $Q_t$. 
\begin{figure}
    \centering
    \includegraphics[scale=0.27]{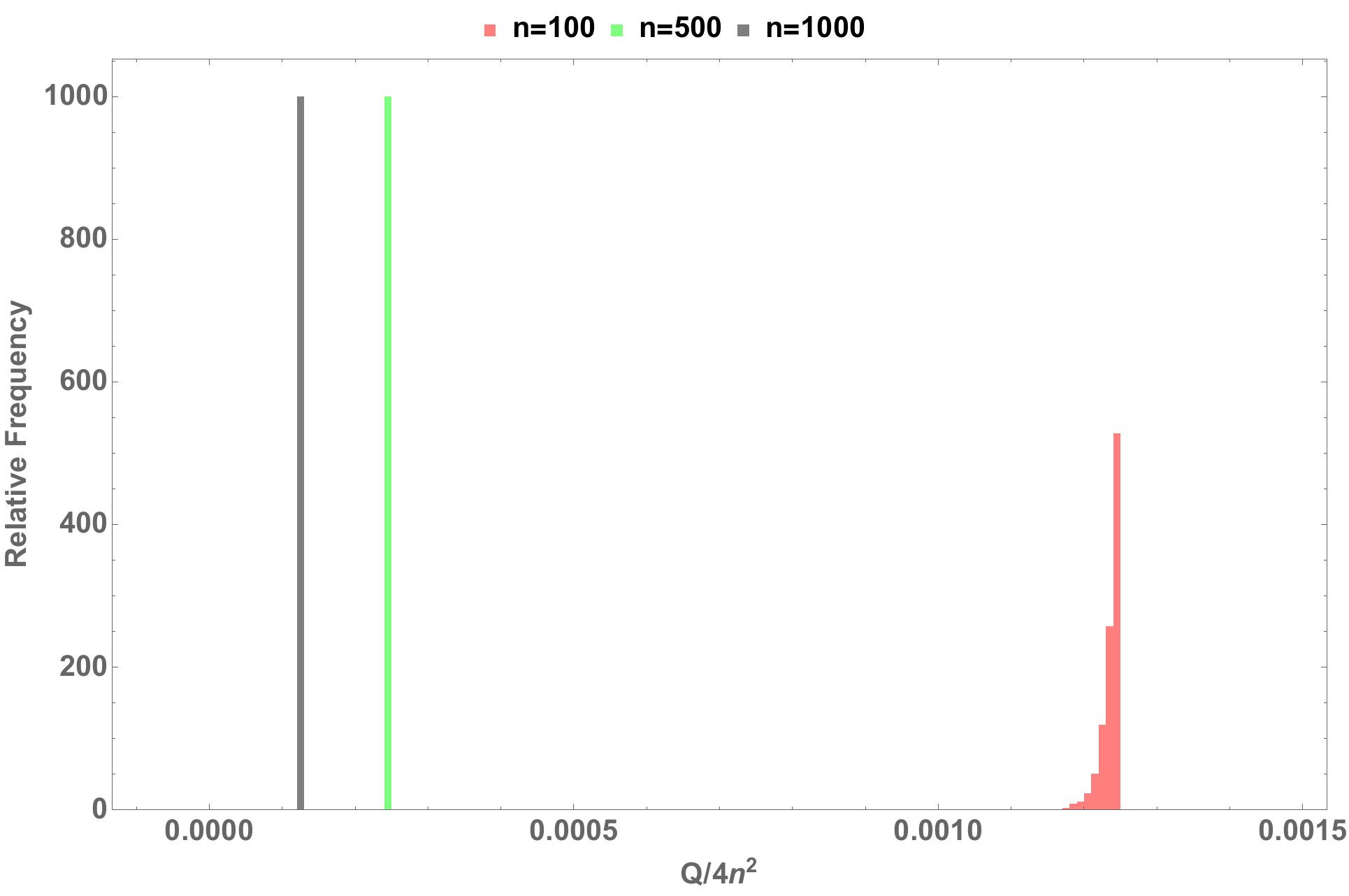}
    \caption{Average of $Q$ over $1000$ samples for random matrices in the Circulant Unitary Ensembles of dimensions $n=10,n=100,500,1000$. The peak of the distribution converges to zero for larger values of $n$.}
    \label{fig:cue}
\end{figure}

How typical is the behavior of a random quantum battery in the large $n$ limit? If there is typicality, an optimal strategy for random quantum batteries would consist in fixing the optimal spectrum of $K$ and then knowing that the other details of the evolution will not matter in the large $n$ limit. To this end, we need to compute the
 fluctuations which is far  more challenging because they involve the fourth tensor power of the unitary representation.  We find that
\ba
\Delta W^2_V= \langle C^2\rangle -\langle C\rangle^2,
\ea
and a  lengthy calculation yields
\ba
\langle C^2\rangle&=&\sum_i\lambda_i \tr \left( \Pi_i(\rho\otimes H_0)^{\otimes 2} (T^{(2)})^{\otimes 2}\right) 
\ea
with $\lambda_i = (\tr \Pi)^{-1}\tr(\Pi_i K^{\otimes 2}\otimes K^{\dagger\otimes 2})$, where $\Pi$'s are the projectors on the irreps of $S_k$, and the index $i$ runs over the five irreducible representations of $S_4$. The details of the calculation are given in Appendix \ref{appendix8a}. 
Let us show that these fluctuations scale like $n^{-2}$. First of all, the expectation values in the above equation can be bound as 
\ba\nonumber 
|\tr [ \Pi_i (\rho\otimes H_0)^{\otimes 2}(T^{(2)})^{\otimes 2}]|&\le& |\tr [(\rho\otimes H_0)^{\otimes 2}]| \nonumber \\
&=& (\tr \rho)^2(\tr H_0)^2\\
& =& (\tr H_0)^2 = O(n^2),
\ea
where the inequality follows from the fact that $\|(T^{(2)})^{\otimes 2}\|\leq 1$.
Putting together all the terms, we find in Appendix \ref{appendix8b} that the fluctuations are upper bounded by 
\ba\nonumber
\Delta W^2_V &\le& O(n^{-4}) M(n)O(n^2) 
\ea
where $M(n)$ is an upper bound to the terms of the form $| \sum_{mnop} e^{i(\theta_{m}+\theta_{p}-\theta_{n}-\theta_{o})} |$. If one chooses spectral properties for $K$ such that $M(n) = O(1)$, then the fluctuations scale like $n^{-2}$ and thus a many-body quantum battery would show exponentially small fluctuations. Moreover, this is the typical case. Indeed, by averaging over CUE to compute $M(n)$, we see in Fig. \ref{fig:cue} that this quantity is concentrated near zero for large $n$. More in depth numerical evidence is provided in Appendix \ref{appendix8b}, where we analyze numerically every single term which contributes to the fluctuations, showing that indeed every single term converges to zero for large $n$'s.

This represents the first main result of this paper: \textit{random quantum batteries show typicality in allowing a work extraction given by the difference in energy between initial state and completely mixed state}, amplified (or attenued) by the form factor $1+Q_t/n^2$. By thus choosing a suitable $V_0$, one can obtain with probability almost one the desired behavior for work extraction in the sense of the Haar measure on $GV_0G^\dagger$.

%%%%%%%RRR
\section{Applications}
\subsection{Jaynes-Cummings model.} The specific behavior of $Q$ determines whether the quantum advantage in a random battery is washed out or not in the large $n$ limit. We
now apply these findings in the case of an exactly solvable model and study the behaviour of $Q$. We consider a two-level system in an optical trap described by the Jaynes-Cummings model \cite{jcm}. In the rotating wave  approximation only two adjacent modes at time $(n,n+1)$ of the electromagnetic field couple with the two level system (details  provided Appendix \ref{appendix8c}). For this calculation, we assume that the atom couples with a finite set of modes of electromagnetic field, which we truncate at a number $n=2 R$, where $R$ is a truncation of the number of modes of the electric field. At the end of the calculation we will send $R\rightarrow \infty$. 

The Hamiltonian reads
\begin{eqnarray}
    H&=&\omega(t) a^\dagger a+ \frac{\Omega(t)}{2} \sigma_z+g (t) (a\sigma_++ a^\dagger \sigma_-) \nonumber\\
    &\equiv& H_0+V(t)
\end{eqnarray}
where we define $\Delta(t)=\Omega(t)-\omega(t)$, and we assume $g(t)=g_0 e^{Mt}$.
%%%%%%%%%%%%%%%%%%%%%%%%%%%%%%
%FIGURE
%%%%%%%%%%%%%%%%%%%%%%%%%%%%%%

\begin{figure}
    \centering
    \includegraphics[scale=0.34]{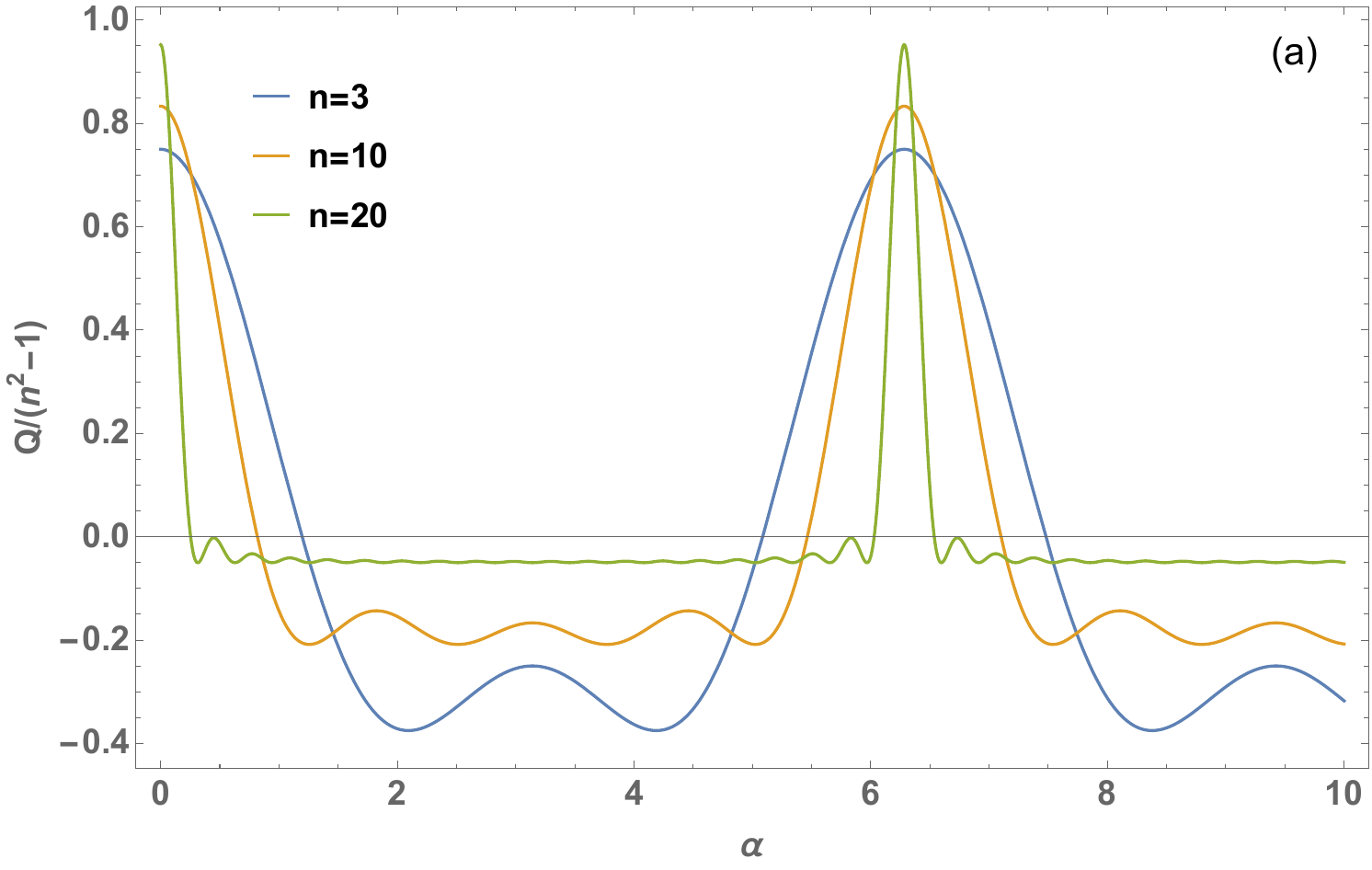}\\
    \includegraphics[scale=0.34]{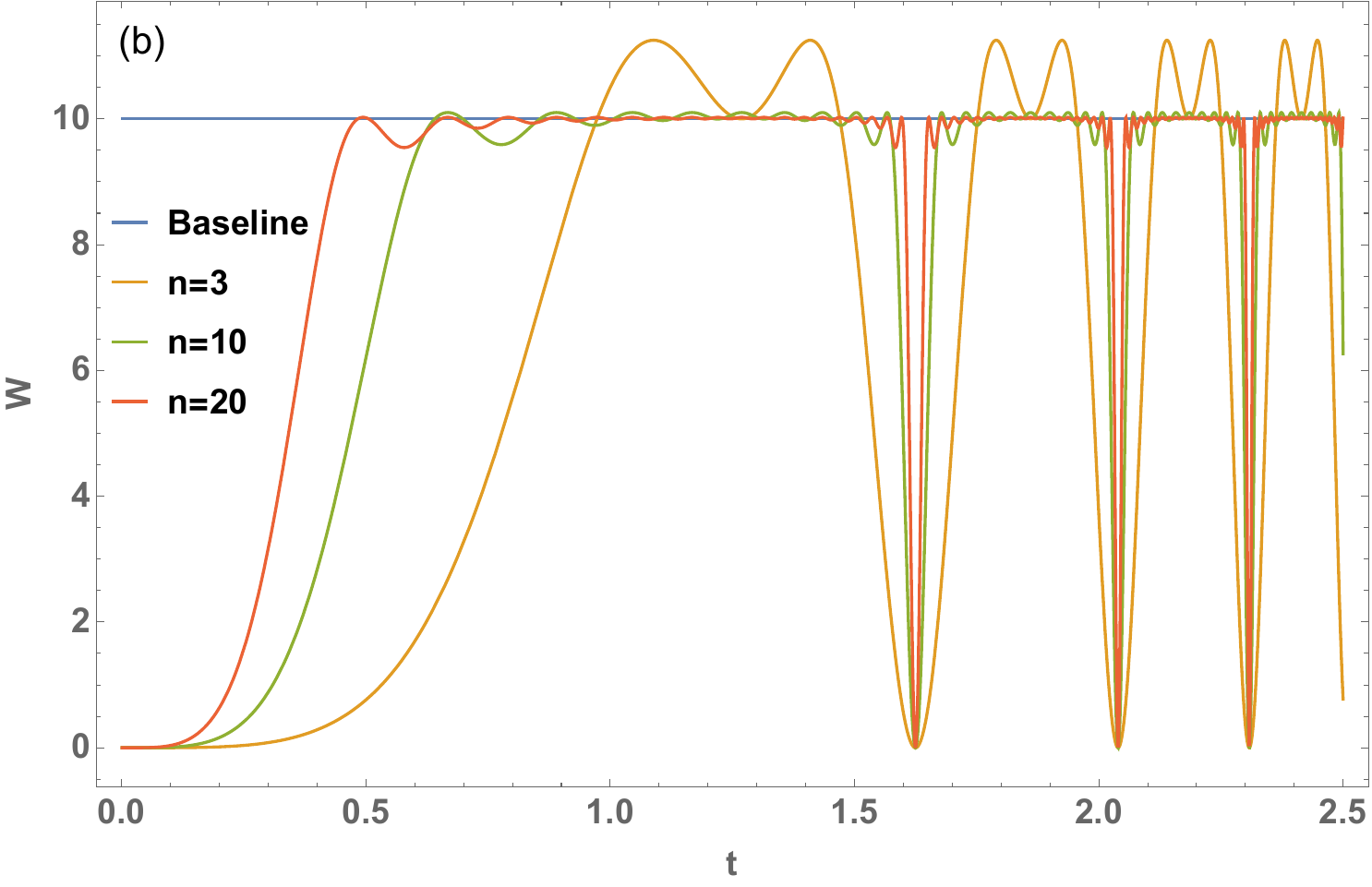}
    \caption{Average work extraction for a Random Quantum Battery made by an optical trap described by the Jaynes-Cummings model. Figure (a): The function $Q(\alpha)$ as function of $\alpha$ for $n=2,10,20$. The maximum value of this function is $0.5$. As the size increases, revivals become more peaked. Figure (b): Work for the Jaynes-Cummings model as a function of time for $\rho=0.5$ for $n=2,10,20$ and $\text{Tr}(H_0)=90 *n $ and $E_0=100$. The baseline represents the work extracted by a battery that brings the system in the completely mixed state.}
    \label{fig:q}
\end{figure}

For this model, we find the eigenvalues $\exp (i\theta_k)$ exactly and use them to evaluate Eq. (\ref{WV2}). Following the calculation in \ref{appendix8c}, we get $\theta_k-\theta_m=g_0^2 (k-m) \left(\frac{e^{M t}-e^{M t_0}}{M^2} \right)^2\equiv(k-m)\alpha_t$, where $M$ is a constant defined as $\frac{\Delta(t)}{\Delta(t^\prime)}=\frac{g(t)}{g(t^\prime)}=e^{M(t-t^\prime)}$. We then obtain the average work Eq.(\ref{avwork})
where, as seen above, the function $Q(\alpha_t)$ is a sum of trigonometric functions whose complete expression is given in Appendix \ref{appendix8c},  Eq.(\ref{thankfully}). 
We note that when $Q>0$, effectively the system extracts more work than the classical counterpark. In this sense, Fig. \ref{fig:q} (a) shows that there can be a quantum advantage in a specific model.

In Fig. \ref{fig:q} we plot the time evolution of the extracted work from the random Jaynes-Cummings battery averaged over $V$. As we can see, for most times the quantum efficiency gets washed out. For small $n$, at specific revival times given by inverting Eq. (\ref{thankfully}), the value of $Q$ becomes of order one, and thus providing a non-vanishing quantum efficiency. This is at the price of performing much worse at different times. One can design a quantum battery by an array of many random nano-batteries of small $n$ and evolve to the revival time where the work extracted goes above that corresponding to the maximally mixed state \cite{pollini,battp, ModiPRL2017}. The fact that non-vanishing $Q$ is obtained as revivals in Eq.(\ref{WV2}) is a sign that this amplification comes from the constructive interference coming from the complex eigenvalues of $K$ and therefore of its quantum nature. On the other hand, for large $n$, the system almost always behaves like in the limit of the battery that completely mixes the state, though one has obtained this performance with limited, realistic resources that do not require to bring the system at infinite temperature.

\subsection{Time dependent perturbation theory}
In the case of the Jaynes-Cummings model we could solve for the time evolution exactly, finding expressions for the average work and its fluctuations via perturbation theory. We make use of the Dyson series for the evolution operator in the interaction picture, namely 
$U_I(t)=\mathcal T \sum_{n=0}^\infty \frac{(-i)^n}{n!} (\int_0^t dt^\prime V_I(t^\prime))^n$. We consider perturbations up to the second order in the Dyson series,
%$U_2= G^\dagger \Big(\mathbb I -i \int^t_{t_0}  V_0(t^\prime)  dt^\prime -\frac{1}{2}   \int^t_{t_0}\int^t_{t_0}:  V_0(t^\prime) V_0(t^{\prime \prime}):  dt^\prime dt^{\prime \prime}\Big)G$. 
and at this point we can average over $G$. Define the operator $A=\int^t_{t_0}  V_0(t^\prime)  dt^\prime$. Again we need the fluctuations of $A$ in the completely mixed state, namely $n^2\Delta A^2= n\ \tr A^2-(\tr A)^2$.

Averaging over $G$ requires a lengthy calculation (see Appendix \ref{appendix8d}) yielding 
\ba
\langle W(t)\rangle_V =\frac{ n^2\Delta A^2 }{n^2 - 1}  \left(E_0-\frac{\tr(H_0)}{n}\right)
\ea
The second term is the difference between the initial energy and the energy in the completely mixed state. %Notice that one cannot extract any work indeed from the completely mixed state. \LPcomm{not sure I understand this last comment}

\begin{figure}
    \centering
    \includegraphics[scale=0.34]{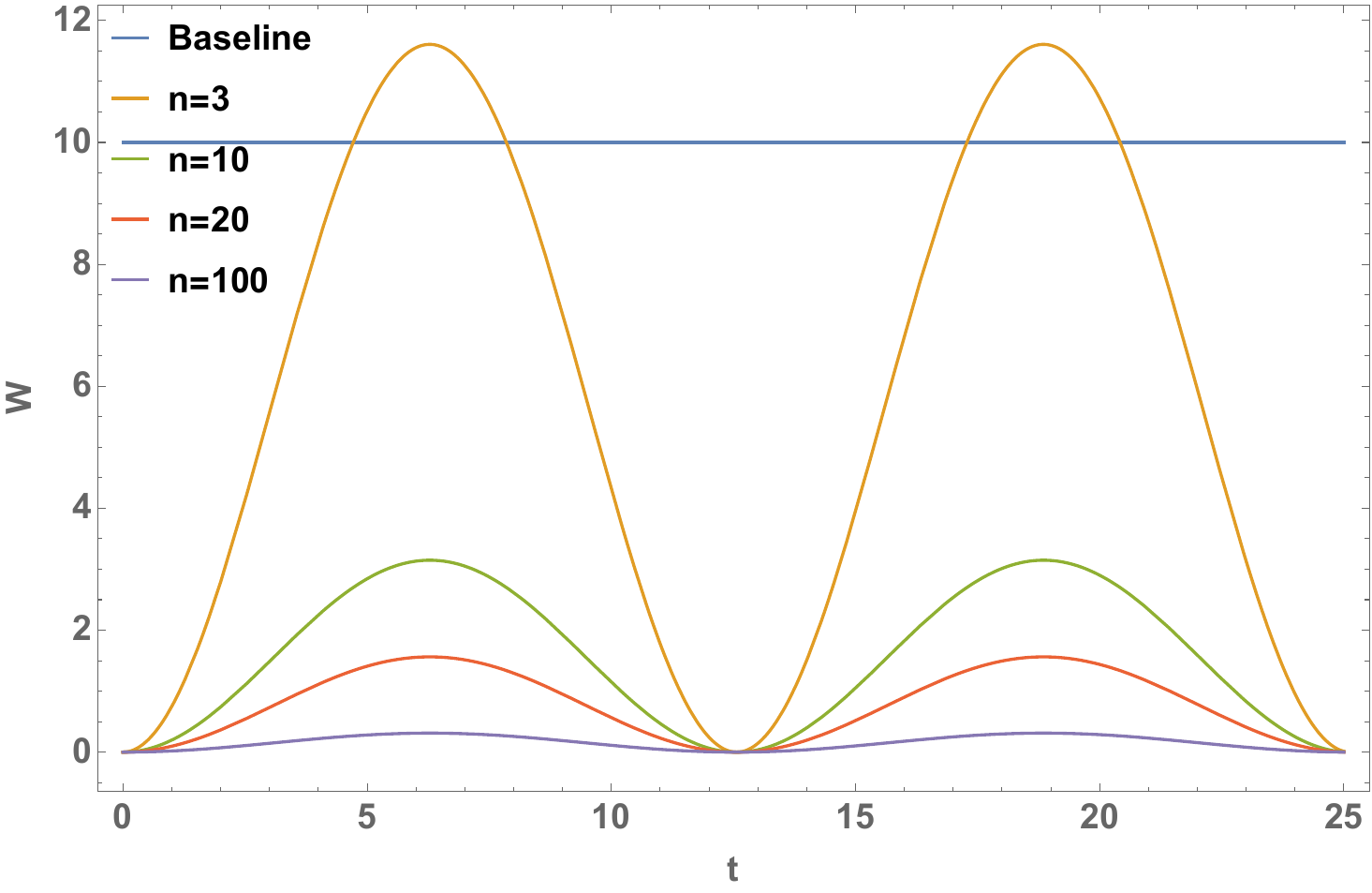}
    \caption{Average work from eqn. (\ref{eq:avw}) for $n=3,10,20,100$ and $\omega=0.5$, against the baseline work $E_0-\frac{\text{Tr}(H_0)}{n}$, with $\text{Tr}(H_0)=90n $ and $E_0=100$.}
    \label{ptheory}
\end{figure}

As an example consider the case of an exactly solvable Hamiltonian $H_0$ subject to the Harmonic perturbation $V(t) =\hat V_0 e^{i \omega t}+\hat V_0^\dagger e^{-i \omega t}$.  Let $\lambda_k$ be the eigenvalues of $\hat V_0$. Averaging over $V$  results in
\begin{eqnarray}
   \langle W(t)\rangle_V &=&\frac{2 f^2(t,\omega)}{(n^2-1)} \Big(\sum_{k,k^\prime} \text{Re}\left(\lambda_k e^{i \frac{t+t_0}{2} \omega}\right)\text{Re}\left(\lambda_{k^\prime} e^{i \frac{t+t_0}{2} \omega}\right) \nonumber \\
    &-&2n \sum_k\text{Re}\left(\lambda_k^2 e^{i \omega (t_0+t)}\right)+2n \sum_k \sigma_k   \Big),
\end{eqnarray}
where we have defined  $f(t,\omega)=2{\sin(\frac{t-t_0}{2} \omega ) }/{\omega}$ and $\lambda$'s are the eigenvalues of $\hat V$. As one can see, the average work decreases with $n$.  We plot $\langle W(t)\rangle_V$ in  Fig.\ref{ptheory}. In this model it is easy to find the revival times at which the quantum efficiency is maintained also for larger values of $n$. 
One can indeed show (see Appendix \ref{appendix8d}) that  the work performed by a random harmonic perturbation of the form $ 2\hat V  \cos(\omega t)$ has always a single maximum at $t_k=(2k+1) \frac{\pi}{\omega}$ on average.

\subsection{Adiabatic Quantum Batteries}\label{adiabscen}
Now let us consider the case of a quantum battery performing an adiabatic evolution connecting the two Hamiltonians $H_0$ and $H_1$ and the two respective equilibrium states $\rho_0,\rho_1$, e.g., two eigenstates or Gibbs states for $H_0,H_1$ (but also thermal or more general mixed equilibrium states). Adiabatic evolution as a method to perform quantum computation\cite{aqc} or quantum control has been long an important tool in quantum information processing, see, e.g., \cite{quiroz}. Adiabatic evolution to perform work extraction was studied in \cite{AdC2014}. A model for an adiabatic quantum battery based on a three-level system was studied in \cite{Zinner}. In this section, we deal with general adiabatic quantum batteries in which the adiabatic drive is rotated in a random direction as a function of time.  

In general, two Hamiltonians  are adiabatically connectible if and only if they belong to the same connected component of the set of iso-degenerate Hamiltonians \cite{adpower}. 
By denoting $H_\alpha=\sum_{i=1}^R \epsilon_\alpha^i
\Pi_\alpha^i \,(\alpha=0,1)$ the spectral resolution of   $H_0$ and $H_1$, and ordering 
 their eigenvalues in ascending order i.e., $\epsilon_\alpha^1<...<\epsilon_\alpha^R.$ We define  the vectors $D_\alpha:=(\rm{tr} \Pi_\alpha^1,\ldots,\Pi_\alpha^R)\equiv (d^1_\alpha\ldots d^R_\alpha)$, with $\sum_k d_\alpha^k = n$. The Hamiltonians  $H_0$ and $H_1$ belong  to the same connected component of the set of iso-degenerate hamiltonians
iff  $D_0=D_1$. So, speaking of adiabatically connected Hamiltonians, we can drop the index $\alpha$.
Let us now introduce the functions $\epsilon^i\colon [0,\,1]\mapsto \R$ 
 such that $\epsilon^i(0)=\epsilon_0^i,$ and $\epsilon_i(1)=\epsilon_1^i\, ((i=1,\ldots,R)$ obeying the
 no-crossing condition $\epsilon^{i+1}(t)>\epsilon^{i}(t)\,
(i=1,\ldots,R-1)$. A continuous family of Hamiltonians connecting $H_0,H_1$ has then the form $H(t)=\sum_{i=1}^R \epsilon^i(t) U_t \Pi_0^i U_t^\dagger,$
where the continuous unitary family $\{U_t\}_{t=0}^1$ is such that $U_0=\openone$ and $U_1=U$.
The work extracted after the adiabatic evolution thus reads
\begin{eqnarray}
    W
    &=&\tr(\rho_0 H_0)-\tr(\rho_1 H_0) \nonumber \\
    &=& \sum_{i=1}^R\tr(p_i (\Pi_0^i-\Pi_1^i) H_0) \nonumber \\
    &=& \sum_{i,j=1}^Rp_i \epsilon^j_0\tr( (\Pi_0^i-\Pi_1^i) \Pi_0^j)
\end{eqnarray}
 because the populations in the $i-$th subspace are conserved by the adiabatic evolution. 
We now have $\Pi_\alpha ^i \Pi_\beta ^j= \delta^{ij}$ if $\alpha=\beta$, but otherwise they are not necessarily orthogonal. We see that the work depends on the choice of $U$ as
\begin{eqnarray}
    W_U&=&\sum_{ij} p_i \epsilon^j_0 \left( \tr(\Pi^i_0\Pi^j_0)-\tr(\Pi^i_0\Pi^j_1) \right) \nonumber \\
    &=& \sum_{ij} p_i \epsilon^j_0 \left( d_i \delta_{ij}-\tr(\Pi^i_0 U \Pi^j_0 U^\dagger) \right)
\end{eqnarray}
We can now perform the average over the unitary transformation $U$. We easily obtain
\begin{eqnarray}
\langle W\rangle_{ad}&=&\sum_{ij} p_i \epsilon^j_0 \left( d_i \delta_{ij}-\tr(\Pi^i_0  \frac{d_j \mathbb I}{n}) \right) \nonumber\\
    &=&\sum_{ij} p_i \epsilon^j_0 \left( d_i \delta_{ij}-  \frac{ d_i d_j }{n} \right) \\
    &=& E_0 - \sum_{ij} p_i \epsilon^j_0  \frac{ d_i d_j }{n} 
\end{eqnarray}
To understand the role of the degeneracies, let us  consider the case of a non degenerate Hamiltonian, so that $d_i=1$ for all $i$. We obtain  $\langle W\rangle_{ad}= E_0-\tr H_0/n$, which again is the difference between the initial energy and the energy of the completely mixed state and thus the quantum efficiency is washed out (see \cite{tb}).  More generally, as we show in Appendix \ref{appendix8f}, we find an upper bound on the adiabatic work given by 
\ba
\langle W\rangle_{ad}&\leq& E_0 (1+c)-\frac{\tr(H_0)}{n}\\
c&=&\frac{\sum_i d_i^2-n}{n}
\ea
so that potentially random adiabatic quantum batteries could give an advantage over classical devices as well (even at infinite temperature), as $c\geq 0$. 

Let us now look at the fluctuations  $\Delta W^2_{ad}$. The calculation involves averaging the square of the work and thus the order two tensored representation of the Unitary group. This is also a lengthy calculation, whose details are given in \ref{appendix8f}. We obtain
\begin{eqnarray}\nonumber
\Delta W^2_{ad}&=&\sum_{i,j,k,l} p_i \epsilon^j_0 p_k \epsilon^l_0 
\times( \frac{d_i d_j d_k d_l}{n^2-1}   -\frac{d_i d_k d_l \delta _{lj}}{n
  (n^2-1)}\\
  &-&\frac{d_j d_k d_l \delta _{ki}}{n
   (n^2-1)}  +\frac{d_k d_l \delta _{ki} \delta _{lj}}{n^2-1}-\frac{d_i d_j d_k d_l}{n^2} ) 
   \end{eqnarray}
For $n\gg 1$, the terms of order $1/n^3$ go to zero faster than $1/n^2$, and we obtain
\begin{eqnarray}
\Delta W^2_{ad}&\underbrace{=}_{n\gg1 }&\frac{1}{n^2} \sum_{i,j,k,l} p_i \epsilon^j_0 p_k \epsilon^l_0 (d_k d_l \delta _{ki} \delta _{lj}) \nonumber \\
&=&\frac{\text{Tr}(H_0 \rho_0)^2}{n^2} =\frac{ E_0^2}{n^2}
\end{eqnarray}
which shows that random adiabatic quantum batteries feature typicality. Fluctuations during adiabatic driving were studied in a different context also in \cite{Funo17}.

\section{Conclusions and Outlook} In this paper we provided a notion of quantum random batteries by means of Haar averaging initial states, Energy measurement Hamiltonian, and the time-dependent driving of the quantum battery. This method allows to study large classes of systems, including not-exactly solvable systems or adiabatic quantum batteries. The average work and fluctuations are systematically studied; we find that quantum batteries exhibit typical behavior in the large $n$ limit given the spectral properties of the driving system. On average, the work extracted is found to be typically equal to the difference between the energy of the initial state and that of the completely mixed state, amplified by a quantum efficiency factor $1+Q_t/n^2$ that only depends on the spectrum of the driving Hamiltonian. Quantum efficiency is not washed out at specific revival times for small systems. Our method allows for the computation of $Q_t$ in perturbation theory, therefore allowing for the treatment of realistic systems. We have also treated the case of random adiabatic quantum batteries, finding that amplification is lost for a non-degenerate Hamiltonian. 

In perspective, our results put forward  several  questions that we would like to investigate in the immediate future. We have shown that for small systems there are revival times in which quantum coherence builds up and gives a quantum advantage. Typically, this is not the case for large $n$. However, it is an open problem whether there are random quantum batteries whose spectral properties allow for the build-up of coherence that outperforms the classical case. Conversely, showing the impossibility of such quantum amplification for large $n$ would be an important result in quantum thermodynamics. This is a problem which we plan to explore  in the near future in a realistic model. A second question relating to the effect of quantum coherence also arises. As we have seen, the extracted work can be related to the coherence of the initial state in {\em two} different bases, or of the operator $U_t$ in two different bases. This suggests that there is a non trivial interplay between coherence and work that involves more than one basis \cite{zanardicoh}. Also, the lack of commutativity between the initial state and the evolution operator or the measuring Hamiltonian and the evolution operator take the form of out of time order correlators. It would then be interesting to explore the connection between fast decays of these quantities, chaos, scrambling, and work statistics. One very intriguing insight comes from the fact that the narrowing of fluctuations does shrink the quantum efficiency but at specific revival times. These revival times correspond to spectral properties of the time evolution operator and one would be interested in understanding the connection between quantum efficiency of random quantum batteries and the integrability or chaotic behavior of the Hamiltonian. 
Using tools from local Haar averaging \cite{hammaavg}, we can explore whether the  efficiency in a battery with a microscopic local drive is  influenced by quantum chaos or integrability. The optimization of the path in a adiabatic quantum algorithm is related to the brachistochrone or geodesics in the space of the ground state manifold \cite{brac}. It would be very interesting to see if optimal paths correspond to bounds given by quantum thermodynamics.  
Finally, it would be important to generalize these results to the case of open quantum systems.

{\em Acknowledgments.---}  
The work of FC was carried out under the auspices of the NNSA of the U.S. DoE at LANL under Contract No. DE-AC52-06NA25396. FC was also financed via DOE-ER grants PRD20170660 and PRD20190195. LPGP also acknowledges partial support by DoE ASCR Quantum Testbed Pathfinder program (award No. DE- sc0019040), ARO MURI, NSF PFCQC program, ARL CDQI, AFOSR, DoE BES QIS program (award No. DE-sc0019449), and NSF PFC at JQI.  
A.H. wants to thank Robert Whitney for insightful conversations at LPMMC, Grenoble, France.

%%%%%%%%%%%%%%%%%%%%%%%%%%%%%%%%%%%%%%%%%%%%%%%%%
%%%%%%%%%%%%%%%%%%%%%%%%%%%%%%%%%%%%%%%%%%%%%%%%%
%%%%%%%%%%%%%%%%%%%%%%%%%%%%%%%%%%%%%%%%%%%%%%%%%
%%%%%%%%%%%%%%%%%%%%%%%%%%%%%%%%%%%%%%%%%%%%%%%%%
%%%%%%%%%%%%%%%%%%%%%%%%%%%%%%%%%%%%%%%%%%%%%%%%%
%%%%%%%%%%%%%%%%%%%%%%%%%%%%%%%%%%%%%%%%%%%%%%%%%
%SUPPLEMENTAL MATERIAL
%%%%%%%%%%%%%%%%%%%%%%%%%%%%%%%%%%%%%%%%%%%%%%%%%
%%%%%%%%%%%%%%%%%%%%%%%%%%%%%%%%%%%%%%%%%%%%%%%%%
%%%%%%%%%%%%%%%%%%%%%%%%%%%%%%%%%%%%%%%%%%%%%%%%%
%%%%%%%%%%%%%%%%%%%%%%%%%%%%%%%%%%%%%%%%%%%%%%%%%
%%%%%%%%%%%%%%%%%%%%%%%%%%%%%%%%%%%%%%%%%%%%%%%%%
%%%%%%%%%%%%%%%%%%%%%%%%%%%%%%%%%%%%%%%%%%%%%%%%%

%\appendix
\clearpage

{
\onecolumngrid 
\section{Appendix}
\subsection{Calculation of $\Delta W^2_\rho$ }\label{appendix1}
Note that for any operator $A$, $\text{Tr}(A)^2 = \tr A \cdot \tr A = \tr (A\otimes A)$. We thus see that $\Delta W^2 = \langle W^2\rangle =  \langle \tr(\rho\delta H_0 )^{\otimes 2}\rangle = \tr\left[ \langle\rho^{\otimes 2}\rangle \delta H^{\otimes 2}\right]$. The average on the tensored representation $G^{\otimes 2}\rho^{\otimes 2}G^{\dagger\otimes 2}$ is also well known\cite{hammaavg,hammaavg2} and is the linear combination on the irreps of $S_2$ given by $ \langle\rho^{\otimes 2}\rangle  = \sum_\pm \lambda_\pm \Pi_\pm$ with $\lambda_\pm = \tr(\Pi_\pm \rho^{\otimes 2})/ \tr\Pi_\pm$ and $\Pi_\pm = (\bbbone^{\otimes 2} + T^{(2)})/2$ where $T^{(2)})$ is the order two permutation (`swap') operator on $\mathcal H ^{\otimes 2}$. 
 Thus we obtain 
\ba\nonumber
\Delta W^2_\rho &=& \sum_\pm \lambda_\pm \tr(\Pi_\pm\delta H^{\otimes 2})\\
\nonumber
& =&\frac{ \left( (\lambda_+ +\lambda_-) \tr \delta H^{\otimes 2} + (\lambda_+ -\lambda_-) \tr( T^{(2)} \delta H^{\otimes 2})\right)}{2}\\
&=& \frac{  (\lambda_+ +\lambda_-) (\tr \delta H)^2 + \frac{1}{2}(\lambda_+ -\lambda_-)\tr \delta H^2}{2}.
\ea
Using the fact that $\tr\delta H=0$ and finally we obtain
\be
\Delta W^2_\rho = \frac{n\tr\rho^2-1}{n(n^2-1)}\tr \delta H^2 = 2\frac{n\tr\rho^2-1}{n(n^2-1)}\tr H_0^2,
\ee
which is the result we present in the paper. 

\subsection{Work fluctuations averaging on $H_0$}\label{appendix2}
Let us define $R=\rho-\rho_t$. We consider the fluctuations on the work via the averaging on the operator $H_0$. We have
\ba\nonumber
\Delta W^2 _{H_0}&=&  \langle W^2\rangle = \tr\left[ R^{\otimes 2}  \sum_\pm \lambda_\pm \Pi_\pm\right] \\
\nonumber 
&=&     
\frac{1}{2}\sum_\pm\lambda_\pm \tr\left[ R^{\otimes 2} \pm  T^{(2)} R^{\otimes 2}\right]
\\
\nonumber
&=&    \frac{1}{2}\sum_\pm\lambda_\pm \left[ (\tr R )^2 \pm\tr R^2\right] \\
\nonumber
&=&  \frac{1}{2}(\lambda_+-\lambda_-)  \tr R^2 
\ea
where now the coefficients of the projectors are $\lambda_\pm= \tr(\Pi_\pm H_0^{\otimes 2})/ \tr\Pi_\pm$. Direct calculation gives, defining $a\equiv (\tr H_0)^2$ and $b\equiv \tr H_0^2$, 
\ba
\frac{1}{2}(\lambda_+-\lambda_-) &=& \frac{1}{n(n+1)} (a+b) -\frac{1}{n(n-1)} (a-b) \\
&=& \left[ \frac{1}{n(n+1)} -\frac{1}{n(n-1)} \right] a +  \left[ \frac{1}{n(n+1)} -\frac{1}{n(n-1)} \right]b \\
&=& \frac{2}{n(n^2-1)}(nb-a) 
\ea
The work fluctuations can thus be written as 
\ba\nonumber
\Delta W^2 _{H_0} =  \frac{2}{n(n^2-1) }\left(n\tr H^2_0 - (\tr H_0)^2\right)\tr R^2
\ea
Now, consider the fluctuations $\Delta H_0^2$ of the eigenvalues of the Hamiltonian $H_0$, namely the fluctuations of $H_0$ in the completely mixed state $\bbbone/n$. We have 
\ba
\Delta H_0^2 = \frac{1}{n} \tr H^2_0 -\frac{1}{n^2}(\tr H_0)^2
\ea
we then obtain
\ba\nonumber
\Delta W^2 _{H_0} =   \frac{2}{n(n^2-1) }n^2\Delta H_0^2 \tr R^2
\ea

%%%%
\ignore{
\be
\lambda_\pm = \frac{2}{n(n\pm 1)}\left((\tr H_0)^2\pm\tr H_0^2 \right) 
\ee
and 
\ba
\lambda_+ \pm \lambda_-&=&\frac{2}{n(n+ 1)} \left(\tr(H_0)^2+ \tr(H_0^2)  \right)\\
&\pm& \frac{2}{n(n- 1)} \left(\tr(H_0)^2- \tr(H_0^2)  \right) \nonumber \\
     &=& \tr(H_0)^2 (\frac{2}{n(n+ 1)}\pm \frac{2}{n(n- 1)} ) \nonumber \\
     &+&\tr(H_0^2)  (\frac{2}{n(n+ 1)}\mp \frac{2}{n(n- 1)} )
\ea
and since
\begin{eqnarray}
  &&  \frac{1}{n(n+1)}+ \frac{1}{n(n-1)}=\frac{2}{n^2-1}\\
     && \frac{1}{n(n+1)}- \frac{1}{n(n-1)}=\frac{-2}{n(n^2-1)},
\end{eqnarray}
we obtain
\ba
\frac{\lambda_+-\lambda_-}{2} = \frac{2}{n^2-1} \left[ (\tr H_0)^2- \frac{\tr H_0^2}{n}\right]
\ea
and thus
\ba\nonumber
\Delta W^2 &=& \frac{2}{n^2-1} \left[ (\tr H_0)^2- \frac{\tr H_0^2}{n}\right] \tr\omega^2\\
&=& \frac{4}{n^2-1} \left[ (\tr H_0)^2- \frac{\tr H_0^2}{n}\right] \tr(\rho^2-\rho\rho_t).
\ea
}
and thus finally
\ba\nonumber
\Delta W^2_{H_0}  &=&  \langle W^2\rangle_{H_0} = \tr\left[ (\rho-\rho_t)^{\otimes 2}  \sum_\pm \lambda_\pm \Pi_\pm\right] \\
&=&  \frac{2n}{n^2-1 }\Delta H_0^2\tr(\rho^2-\rho\rho_t)
\ea
which is the result we report in the paper.
\subsection{Traces of $K$}\label{appendix3}
A direct calculation of the coefficients yields
\begin{eqnarray}
    \lambda_+&=&\frac{\tr(K \otimes K^\dagger \Pi_+)}{\tr\Pi_+}=\frac{2}{n(n+1)} \frac{\tr K \tr K^\dagger+\tr K K^\dagger }{2} \nonumber \\
    \lambda_-&=&\frac{\tr(U_0 \otimes U_0^\dagger\Pi_-)}{\tr\Pi_-}=\frac{2}{n(n-1)} \frac{\tr K\tr K^\dagger-\tr KK^\dagger}{2} \nonumber 
\end{eqnarray}
Moreover, we use that
\begin{equation}
  \lambda_+ \Pi_++\lambda_- \Pi_-=\frac{\lambda_++\lambda_-}{2} \mathbb{I}+\frac{\lambda_+-\lambda_-}{2} T
\end{equation}
We now see that, defining
\begin{eqnarray}
    a&=&\tr K \tr K^\dagger=|\sum_i e^{i \theta_i} |^2=2 \sum_{j\neq k} \cos(\theta_j-\theta_k)+ n \nonumber \\
    b&=&\tr KK^\dagger=n
\end{eqnarray}
and thus
\begin{eqnarray}
\lambda_+=\frac{1}{n} \frac{a+b}{n+1},\ \ \ \ \lambda_{-}=\frac{1}{n} \frac{a-b}{n-1}.
\end{eqnarray}
Using the relationships
\begin{eqnarray}
    \frac{1}{2n} \left(\frac{a+b}{ (n+1)}+\frac{a-b}{(n-1) }\right)&=&\frac{an-b}{n^3-n} \nonumber \\
    \frac{1}{2n} \left(\frac{a+b}{ (n+1)}-\frac{a-b}{(n-1) }\right)&=&\frac{bn-a}{n^3-n}
    \label{eq:eigs}
\end{eqnarray}
 we get
\begin{eqnarray}
\nonumber
\frac{\lambda_++\lambda_-}{2}&=& \frac{n(2 \sum_{j\neq k} \cos(\theta_j-\theta_k)+ n)-n}{n^3-n}\\
\nonumber &=&\frac{2 \sum_{j\neq k} \cos(\theta_j-\theta_k)+ n-1}{n^2-1} \nonumber \\
\frac{\lambda_+-\lambda_-}{2}&=& \frac{n^2-(2 \sum_{j\neq k} \cos(\theta_j-\theta_k)+ n)}{n(n^2-1)}
\end{eqnarray}

%%%%%%%%%%%%%%%%%%%%

%%%%%%%%%%%%%%%%%%%
\subsection{Calculation of $  \langle W(t)\rangle_V $ and $ \Delta W^2_V$ }\label{appendix4}

The work extracted $W(t)$  reads 
\ba\nonumber
W(t) &=& \tr[\omega H_0] = \tr[\rho H_0] - \tr[\rho_t H_0] \equiv E_0 - \tr[\rho_t H_0] \\
&=& E_0 - \tr [U_I\rho U_I^\dagger H_0]\equiv E_0- C
\ea
We can write the above expression as
\ba
W(t) &=& E_0 -\tr\left[  U_I \rho\otimes U_I^\dagger H_0 T^{(2)}  \right] \\
&=& E_0- \tr\left[(\rho \otimes H_0) (U_I\otimes U_I^\dagger)  T^{(2)} \right]
\ea
The average work over the noise $G$ can then be computed as 
\ba\nonumber
\langle W(t)\rangle_V &=& E_0 - \tr\left[(\rho \otimes H_0) \langle (U_I\otimes U_I^\dagger)\rangle  T^{(2)} \right] \\
\nonumber
&=& E_0 -  \tr\left[(\rho \otimes H_0) \langle (GKG^\dagger\otimes GK^\dagger G^\dagger)\rangle  T^{(2)} \right]  \\
&=& E_0 -  \tr\left[   (\rho \otimes H_0) \langle G^{\otimes 2} (K\otimes K^\dagger) G^{\dagger\otimes 2}\rangle T^{(2)}  \right] \nonumber \\
\ea
The unitary operator $K=\mathcal T \exp (-i\int^t_0 V(s) ds) $ will be diagonalized in the form $K = \sum_{k} \exp({i\theta_k})|k\rangle\langle k|$.

Using the usual technique, we find $ \langle G^{\otimes 2} (K\otimes K^\dagger) G^{\dagger\otimes 2}\rangle  = \sum_\pm \lambda_\pm \Pi_\pm$, 
where now $\lambda_\pm = \tr(\Pi_\pm K\otimes K^{\dagger})/ \tr\Pi_\pm$. Notice that in this setup, already the average work involves the average over the tensored representation of the unitary group. We obtain
\ba
  \langle W(t)\rangle_V &=&E_0-\tr\left(( \frac{\lambda_++\lambda_-}{2} \mathbb{I}+ \frac{\lambda_+-\lambda_-}{2} T)T^{(2)}(\rho_0\otimes H_0)  \right) \nonumber \\
    &=&E_0-\tr\left(( \frac{\lambda_++\lambda_-}{2} T^{(2)}+ \frac{\lambda_+-\lambda_-}{2} \mathbb{I})(\rho_0\otimes H_0)  \right) \nonumber \\
    &=&E_0(1-\frac{\lambda_++\lambda_{-}}{2} )-\frac{\lambda_+-\lambda_-}{2} \tr(\rho_0) \tr(H_0) \nonumber \\
    &=&E_0\left(1-\frac{2 \sum_{j\neq k} \cos(\theta_j-\theta_k)+ n-1}{n^2-1}\right)  \nonumber \\
    &-& \frac{n^2-(2 \sum_{j\neq k} \cos(\theta_j-\theta_k)+ n)}{n^2-1} \frac{\tr(H_0)}{n}.
\ea
We finally obtain
\ba\label{avwork_supp}
\langle W(t)\rangle_V = \left[ E_0-\frac{\tr H_0}{n}\right] \left[ \frac{n}{n+1} +\frac{Q}{n^2-1}\right]
\ea
where
\ignore{\begin{eqnarray}
    W_0&=&E_0\left(1-\frac{ 1}{n+1}\right)- \frac{\tr(H_0)}{n+1} \nonumber \\
    W_1&=& \frac{1 }{n^2-1}\left( \frac{\tr(H_0)}{n}+E_0\right)\\
    \nonumber
    Q(\theta_j-\theta_k) &=& 2\sum_{j\neq k} \cos(\theta_j-\theta_k)
\end{eqnarray}
}
In the above equation, $\exp(i\theta_k)$ are the eigenvalues of the evolution operator $K=\mathcal T \exp (-i\int^t_0 V(s) ds) $. All the time dependence of the  is thus contained in the function $Q(\theta_j-\theta_k)$.

The fluctuations are  more challenging because they involve the fourth tensor power of the unitary representation. Let us set out to find them.  We see that
\ba
\Delta W^2_V= \langle C^2\rangle -\langle C\rangle^2
\ea
where $C\equiv \tr [U_I\rho U_I^\dagger H_0]$. 
The relevant object to compute is then
\ba
\langle C^2\rangle&=& \tr\left(  \langle (U_I\rho U^\dagger_I)^{\otimes 2}\rangle H_0^{\otimes 2}\right) \\
&=& \tr\left(  \langle (GKG^\dagger \rho GK^\dagger G)^{\otimes 2}\rangle H_0^{\otimes 2}\right)\\
&=& \tr \left( \langle G^{\otimes 4}(K\otimes K^\dagger) ^{\otimes 2}G^{\dagger\otimes 4})\rangle (\rho\otimes H_0)^{\otimes 2}\right)
\ea
This time, the average reads
\ba
\langle G^{\otimes 4}(K\otimes K^\dagger) ^{\otimes 2}G^{\dagger\otimes 4}\rangle =\sum_i \lambda_i \Pi_i
\ea
with $\lambda_i = (\tr \Pi)^{-1}\tr(\Pi_i K^{\otimes 2}\otimes K^{\dagger\otimes 2})$. Now, the $\Pi_i$ are the projectors onto the irreps of $S_4$. There are five irreducible irreps of $S_4$. In the next subsection we show  an explicit expression of these projectors. A  lengthy calculation yields

\begin{eqnarray}
\langle C^2\rangle&=&\sum_i\lambda_i \tr \left( \Pi_i(\rho\otimes H_0)^{\otimes 2}\right) \\
&=& \frac{Tr(\Pi^{(tr)} (\rho\otimes H_0)^{\otimes 2} )}{Tr(\Pi^{(tr)} )}+\frac{Tr(\Pi^{(sig)} (\rho\otimes H_0)^{\otimes 2} )}{Tr(\Pi^{(sig)} (\rho\otimes H_0)^{\otimes 2})}+\frac{Tr(\Pi^{(st)} (\rho\otimes H_0)^{\otimes 2})}{Tr(\Pi^{(st} )} \nonumber \\
&+&\frac{Tr(\Pi^{(st\otimes sgn)} (\rho\otimes H_0)^{\otimes 2})}{Tr(\Pi^{(st\otimes sgn})}+\frac{Tr(\Pi^{(2D)} (\rho\otimes H_0)^{\otimes 2})}{Tr(\Pi^{(2D)}) }
\end{eqnarray}

\subsection{Irreps of $S_4$}\label{appendixE}
Let us first recall the character table of $S_4$ in Table \ref{tab:s4}.
\begin{table} 
\centering
\begin{tabular}{l*{4}{c}r}
             & e &(12) & (123)& (1234) & (12)(34)  \\
\hline
trivial & 1 & 1 & 1& 1 & 1  \\
sgn            & 1 & -1 & 1 & -1 &  1  \\
st           & 3 & 1 & 0 & -1 &  -1   \\
$st\otimes sgn$    & 3 & -1 & 0 & 1 &  -1   \\
2D  & 2 & 0 &-1 & 0& 2\\
size      & 1 & 6 & 8 & 6 & 3\\
\end{tabular}
\caption{The character table of $S_4$.}
\label{tab:s4}
\end{table}
The last row of Table \ref{tab:s4} gives the size of each conjugacy class in $S_4$. 
Given a permutation $\sigma\in S_4$, we denote by $S(\sigma)$  the representation of $S:S_4\mapsto GL(\mathcal H^{\otimes 4})$ given by 
\be
S (\sigma) = \sum_{ijkl} |\sigma (ijkl)\rangle \langle ijkl|
\ee
By the Schur-Weyl duality the projectors onto its irreps are 
\ba
\Pi^{(r)} = \frac{\chi^{(r)} (e)}{|S_4|}\sum_{\sigma \in S_4} \chi^{(r)} (\sigma) S(\sigma) 
\ea
where $\chi^{(r)}$ is the character of the $(r)$ irrep of $S_4$ and $\chi^{(e)}$ is the dimension of the irrep in $S_4$.

The five projectors are given by:
\ba\nonumber
\Pi^{(tr)} &=&\frac{1}{24} \sum_{S_4} S(\sigma) \\
\nonumber
\Pi^{(sig)} &=& \frac{1}{24} \left[ \sum_{\sigma even} S (\sigma_{ev}) -  \sum_{\sigma odd} S (\sigma_{odd})\right] \\
\nonumber
\Pi^{(st)}&=&\frac{3}{24}[ 3\bbbone+ (S_{(12)}+ \ldots) -(S_{(1234)}+\ldots) \\
\nonumber
&-& (S_{(12)(34)} +\ldots) ]\\
\nonumber
\Pi^{(st\otimes sgn)}&=& \frac{3}{24} [3\bbbone-(S_{(12)}+ \ldots) +(S_{(1234)}+\ldots) \\
\nonumber
&-& (S_{(12)(34)} +\ldots) ] \\
\nonumber
\Pi^{(2D)}&=& \frac{2}{24} \left[ 2\bbbone -(S_{(123)}+\ldots) +2(S_{(12)(34)}+\ldots)\right]
\ea

In the above, the symbol $+\ldots$ denotes a sum over all the members of the conjugacy class. As well known, the five conjugacy classes of $S_4$ are given by their cycle structure of Table \ref{tab:conjc}.
{\begin{table}
\centering
\begin{tabular}{l*{5}{c}r}
e & (..)& (..)(..) &(...) &(....) \\
\hline
e &(12) &(12)(34) &(123) &(1234)\\
&(13)& (13)(24)& (132) &(1342)\\
&(14) &(14)(23)& (124)& (1423)\\
&(23)&  &(142) &(1243)\\
&(24)&  &(134) &(1432)\\
&(34)&  &(143) &(1324)\\
& & & (234) &\\
& &  & (243) &
\end{tabular}
\caption{The conjugacy classes of $S_4$ which we use for the projectors.}
\label{tab:conjc}
\end{table}
}

\section{Work fluctuations via irreps of $S_4$}
\subsection{Main definitions and projectors}\label{appendix8a}
First, let us note that for any operator $A$, we have $|\text{Tr}(A T^{(2),\otimes 2})|\leq |\text{Tr}(A)|$ since $\|T^{(2),\otimes 2}\|\leq 1 $, and since we care about the scaling of the fluctuations we can focus on a simpler calculation which does not involve the swap.  Before we begin the calculation, we start with a few definitions which will be useful in the following:
\begin{eqnarray}
K&=&\sum_k e^{i\theta_{k}}\ket{k}\bra{k},  \nonumber \\
\langle C^{2} \rangle &=&Tr[\langle(U_{I}\rho U_{I}^{\dagger})^{\otimes 2}\rangle H_{0}^{\otimes 2}]=Tr[\langle(GkG^{\dagger}\rho G^{\dagger}K^{\dagger}G)^{\otimes 2}\rangle H_{0}^{\otimes 2}],\nonumber \\
&=&Tr[\langle G^{\otimes 4}(K\otimes K^{\dagger})^{\otimes 2}  G^{\dagger \otimes 4}\rangle (\rho \otimes H_{0})^{\otimes 2)}],\nonumber \\
\langle G^{\otimes 4} (K\otimes K^{\dagger})^{\otimes 2} G^{\dagger \otimes 4} \rangle&=&\sum_{i}\lambda_{i}\Pi_{i},\nonumber \\
\lambda_{i}&=&  (Tr\Pi_{i})^{-1}Tr[\Pi_{i}K^{\dagger 2}\otimes K^{\dagger \otimes 2}],\nonumber \\
(\rho\otimes  H_{0})^{\otimes 2}&=&\sum_{abcdef}\rho_{ab}\epsilon_{c}\rho_{de}\epsilon_{f}\ket{acdf}\bra{bcef},\nonumber \\
\langle C^{2} \rangle &=& \sum_i \lambda_i \tr \left[ \Pi_i (\rho\otimes  H_{0})^{2} \right],\nonumber 
\end{eqnarray}
We then start with the construction of the projectors in a basis, which we take as the computational basis:
\begin{eqnarray}
S_{(1234)}+...&\propto& \sum_{ijkl} \ket{lijk}\bra{ijkl}+\ket{kilj}\bra{ijkl}+\ket{lkij}\bra{ijkl}+\ket{jlik}\bra{ijkl}+\ket{klji}\bra{ijkl}+\ket{jkli}\bra{ijkl} \nonumber
\end{eqnarray}
\begin{eqnarray}
S_{(123)}+...&\propto&\sum_{ijkl} \ket{kijl}\bra{ijkl}+\ket{jkil}\bra{ijkl}+\ket{iljk}\bra{ijkl}+\ket{iklj}\bra{ijkl}+\ket{ljik}\bra{ijkl}+\ket{kjli}\bra{ijkl} \nonumber \\
&\ &\ \  \ \ +\ket{likj}\bra{ijkl}+\ket{jlki}\bra{ijkl}
\end{eqnarray}
\begin{eqnarray}
S_{(1)}&\propto&\sum_{ijkl} \ket{ijkl}\bra{ijkl}
\end{eqnarray}
\begin{eqnarray}
S_{(12)}+...&\propto&\sum_{ijkl} \ket{jikl}\bra{ijkl}+\ket{kjil}\bra{ijkl}+\ket{ljki}\bra{ijkl}+\ket{ikjl}\bra{ijkl}+\ket{ilkj}\bra{ijkl}+\ket{ijlk}\bra{ijkl}
\end{eqnarray}
\begin{eqnarray}
S_{(12)(34)}+...&\propto&\sum_{ijkl} \ket{jilk}\bra{ijkl}+\ket{klij}\bra{ijkl}+\ket{lkji}\bra{ijkl}
\end{eqnarray}
Since we are interested only in the scaling with $n$ of the fluctuations, we focus on the structure of the traces and not on the proportionality constants.
Using the definitions above, the projectors can then be written explicitly in the computational basis.
At this point, we can start the evaluation of the traces. 
First we note that $(\rho\otimes  H_{0})^{2}=\sum_{abcdef}\rho_{ab}\epsilon_{c}\rho_{de}\epsilon_{f}\ket{acdf}\bra{bcef}$.
We then have:
%$Tr[\Pi(\rho\otimes H_{0})^{\otimes2}]$:
} 

\begin{eqnarray}
%24
\text{Tr}[\Pi^{(tr)}(\rho\otimes H_{0})^{\otimes2}] %\text{Tr}[(\rho\otimes H_{0})^{\otimes2}]+Tr[(S_{(12)(34)}+...)(\rho\otimes H_{0})^{\otimes2}]\\
%\nonumber \\
%&+&\text{Tr}[(S_{(123)}+...)(\rho\otimes H_{0})^{\otimes2}]+Tr[(S_{(1234)}+...)(\rho\otimes H_{0})^{\otimes2}]+Tr[(S_{(12)}+...)(\rho\otimes H_{0})^{\otimes2}]\nonumber  \\
%\ea
%\ba
%\nonumber &=&\sum_{ijkl}(\bra{ijkl}(\rho\otimes H_{0})^{\otimes2}\ket{jikl}+\bra{ijkl}(\rho\otimes H_{0})^{\otimes2}\ket{klij}+\bra{ijkl}(\rho\otimes H_{0})^{\otimes2}\ket{lkji})\\
%\nonumber&+&\sum_{ijkl}(\bra{ijkl}(\rho\otimes H_{0})^{\otimes2}\ket{kijl}+\bra{ijkl}(\rho\otimes H_{0})^{\otimes2}\ket{jkil}+\bra{ijkl}(\rho\otimes H_{0})^{\otimes2}\ket{iljk}\\
%\nonumber&+&\bra{ijkl}(\rho\otimes H_{0})^{\otimes2}\ket{iklj}+\bra{ijkl}(\rho\otimes H_{0})^{\otimes2}\ket{ljik}+\bra{ijkl}(\rho\otimes H_{0})^{\otimes2}\ket{jkil}+
%\\ 
%& &\nonumber \bra{ijkl}(\rho\otimes H_{0})^{\otimes2}\ket{likj}+\bra{ijkl}(\rho\otimes H_{0})^{\otimes2}\ket{jlki}) +... \\
%\ea
%\ba
%&=& \sum_{abcdeijkl} \rho_{ab}\epsilon_{c}\rho_{de}\epsilon_{f}\langle ijkl | acdf \rangle \langle bcef | jikl \rangle +\cdots\\ 
&\propto&
%\sum_{abcdef} 
\sum
%\Big(\rho_{af}\epsilon_{a}\rho_{dc}\epsilon_{d}+\rho_{ad}\epsilon_{a}\rho_{dc}\epsilon_{d}+\rho_{af}\epsilon_{d}\rho_{da}\epsilon_{c}+\rho_{ac}\epsilon_{f}\rho_{da}\epsilon_{d}+\rho_{ad}\epsilon_{f}\rho_{da}\epsilon_{a}\\ & &\nonumber+\rho_{ac}\epsilon_{d}\rho_{df}\epsilon_{a}+\rho_{ad}\epsilon_{a}\rho_{dc}\epsilon_{f}+\rho_{ac}\epsilon_{d}\rho_{da}\epsilon_{f}+\rho_{aa}\epsilon_{f}\rho_{dc}\epsilon_{d}+\rho_{aa}\epsilon_{d}\rho_{df}\epsilon_{c}+\rho_{af}\epsilon_{c}\rho_{da}\epsilon_{d}\\ 
%\nonumber &+ &\rho_{af}\epsilon_{a}\rho_{dd}\epsilon_{c}+\rho_{ac}\epsilon_{f}\rho_{dd}\epsilon_{a}+\rho_{aa}\epsilon_{c}\rho_{dd}\epsilon_{f}+\rho_{ac}\epsilon_{a}\rho_{dd}\epsilon_{f}+\rho_{ad}\epsilon_{c}\rho_{da}\epsilon_{f}+\rho_{af}\epsilon_{c}\rho_{dd}\epsilon_{a}+\\ 
%\nonumber & &\rho_{aa}\epsilon_{d}\rho_{dc}\epsilon_{f}+\rho_{aa}\epsilon_{f}\rho_{dd}\epsilon_{c}+\rho_{aa}\epsilon_{c}\rho_{df}\epsilon_{d}+ \rho_{ac}\epsilon_{a}\rho_{df}\epsilon_{d}+\nonumber\rho_{ad}\epsilon_{f}\rho_{da}\epsilon_{c}+\rho_{af}\epsilon_{d}\rho_{dc}\epsilon_{a}\Big)
\Big[ \rho_{aa}\epsilon_{c}\rho_{dd}\epsilon_{f} + \rho_{ad}\epsilon_{a}\rho_{de}\epsilon_{f}+\rho_{ac}\epsilon_{d}\rho_{da}\epsilon_{f}
+\rho_{aa}\epsilon_{f}\rho_{dc}\epsilon_{d}+\rho_{aa}\epsilon_{d}\rho_{df}\epsilon_{c}+\rho_{af}\epsilon_{c}\rho_{da}\epsilon_{d} \nonumber  \\ 
\nonumber& &\ \ \ \ \ \ \ \ \ \ +\rho_{ad}\epsilon_{c}\rho_{df}\epsilon_{a}+\rho_{af}\epsilon_{a}\rho_{dd}\epsilon_{c}+\rho_{ac}\epsilon_{f}\rho_{dd}\epsilon_{a}+\rho_{ac}\epsilon_{a}\rho_{df}\epsilon_{d}+\rho_{ad}\epsilon_{f}\rho_{da}\epsilon_{c}+\rho_{af}\epsilon_{d}\rho_{dc}\epsilon_{a}+\\ 
\nonumber& &\ \ \ \ \ \ \ \ \ \ \rho_{ad}\epsilon_{a}\rho_{df}\epsilon_{c}+\rho_{af}\epsilon_{d}\rho_{da}\epsilon_{c} +\rho_{ac}\epsilon_{f}\rho_{da}\epsilon_{d}+\rho_{ad}\epsilon_{f}\rho_{da}\epsilon_{a}+\rho_{ac}\epsilon_{d}\rho_{df}\epsilon_{a}+\rho_{ac}\epsilon_{a}\rho_{dd}\epsilon_{f}
\\ \nonumber& &\ \ \ \ \ \ \ \ \ \ +\rho_{af}\epsilon_{a}\rho_{dc}\epsilon_{d}+\rho_{ad}\epsilon_{e}\rho_{da}\epsilon_{f}+\rho_{af}\epsilon_{c}\rho_{dd}\epsilon_{a}+\rho_{aa}\epsilon_{d}\rho_{dc}\epsilon_{f}+\rho_{aa}\epsilon_{f}\rho_{dd}\epsilon_{c}+\rho_{aa}\epsilon_{c}\rho_{df}\epsilon_{d}\Big] 
\end{eqnarray}

\begin{eqnarray}
%24
\text{Tr}[\Pi^{(sig)}(\rho\otimes H_{0})^{\otimes2}]&\propto&
\sum\Big(
\rho_{aa}\epsilon_{c}\rho_{dd}\epsilon_{f} + \Big(\rho_{ad}\epsilon_{a}\rho_{de}\epsilon_{f}+\rho_{ac}\epsilon_{d}\rho_{da}\epsilon_{f}
%\\ 
%& &\nonumber 
+\rho_{aa}\epsilon_{f}\rho_{dc}\epsilon_{d}+\rho_{aa}\epsilon_{d}\rho_{df}\epsilon_{c}+\rho_{af}\epsilon_{c}\rho_{da}\epsilon_{d}
\\
& &\nonumber
+\rho_{ad}\epsilon_{c}\rho_{df}\epsilon_{a}+\rho_{af}\epsilon_{a}\rho_{dd}\epsilon_{c}+\rho_{ac}\epsilon_{f}\rho_{dd}\epsilon_{a}
%\\ 
%& &\nonumber 
+\rho_{ac}\epsilon_{a}\rho_{df}\epsilon_{d}+\rho_{ad}\epsilon_{f}\rho_{da}\epsilon_{c}+\rho_{af}\epsilon_{d}\rho_{dc}\epsilon_{a}\Big)
\\ 
& &\nonumber-\Big(\rho_{af}\epsilon_{a}\rho_{dc}\epsilon_{d}+\rho_{ad}\epsilon_{a}\rho_{df}\epsilon_{c}+\rho_{af}\epsilon_{d}\rho_{da}\epsilon_{c}
%\\ 
%& &\nonumber
+\rho_{ac}\epsilon_{f}\rho_{da}\epsilon_{d}+\rho_{ad}\epsilon_{f}\rho_{da}\epsilon_{a}+\rho_{ac}\epsilon_{d}\rho_{df}\epsilon_{a}
\\ 
& &\nonumber+
\rho_{ac}\epsilon_{a}\rho_{dd}\epsilon_{f}+\rho_{ad}\epsilon_{e}\rho_{da}\epsilon_{f}+\rho_{af}\epsilon_{c}\rho_{dd}\epsilon_{a}\Big)
%\\ 
%& &\nonumber
+\rho_{aa}\epsilon_{d}\rho_{dc}\epsilon_{f}+\rho_{aa}\epsilon_{f}\rho_{dd}\epsilon_{c}+\rho_{aa}\epsilon_{c}\rho_{df}\epsilon_{d}\Big)
\end{eqnarray}
\begin{eqnarray}
%\frac{24}{3}
\text{Tr}[\Pi^{(st)}(\rho\otimes H_{0})^{\otimes2}]&\propto&
%&=&3\text{Tr}[(\rho\otimes H_{0})^{\otimes2}]-\text{Tr}[(S_{(12)(34)}+...)(\rho\otimes H_{0})^{\otimes2}] \nonumber \\
%&-&\text{Tr}[(S_{(1234)}+...)(\rho\otimes H_{0})^{\otimes2}]+\text{Tr}[(S_{(12)}+...)(\rho\otimes H_{0})^{\otimes2}] \nonumber \\
%&=&3\text{Tr}^{2}[H_{0}] \nonumber \\
%& & -[\sum_{ijkl}(\bra{ijkl}(\rho\otimes H_{0})^{\otimes2}\ket{jikl}+\bra{ijkl}(\rho\otimes H_{0})^{\otimes2}\ket{klij}+\bra{ijkl}(\rho\otimes H_{0})^{\otimes2}\ket{lkji})]+...\nonumber \\
%&=&3\text{Tr}\Big[(\rho\otimes H_{0})^{\otimes2}\Big] \nonumber \\
%& &-\Big[
%\sum_{abcdef} 
\sum
\Big(\rho_{ac}\epsilon_{a}\rho_{dd}\epsilon_{f}+\rho_{ad}\epsilon_{c}\rho_{da}\epsilon_{f}+\rho_{af}\epsilon_{c}\rho_{dd}\epsilon_{a}\\  &\ &\ \ \ \ \ \ \ \ \ \ \ \ +
\nonumber \rho_{aa}\epsilon_{d}\rho_{dc}\epsilon_{f}+\rho_{aa}\epsilon_{f}\rho_{dd}\epsilon_{c}+\rho_{aa}\epsilon_{c}\rho_{df}\epsilon_{d}\Big)-\Big(\rho_{af}\epsilon_{a}\rho_{dc}\epsilon_{d}+\rho_{ad}\epsilon_{a}\rho_{dc}\epsilon_{d} \\
&\ &\ \ \ \ \ \ \ \ \ \ \ \ +\rho_{af}\epsilon_{d}\rho_{da}\epsilon_{c}+\rho_{ac}\epsilon_{f}\rho_{da}\epsilon_{d}+\rho_{ad}\epsilon_{f}\rho_{da}\epsilon_{a}+\rho_{ac}\epsilon_{d}\rho_{df}\epsilon_{a}\Big) \nonumber \\
&\ &\ \ \ \ \ \ \ \ \ \ \ \ -\Big(\rho_{ac}\epsilon_{a}\rho_{df}\epsilon_{d}+\nonumber\rho_{ad}\epsilon_{f}\rho_{da}\epsilon_{c}+\rho_{af}\epsilon_{d}\rho_{dc}\epsilon_{a}\Big)\Big]
\end{eqnarray}

\begin{eqnarray}
%\frac{24}{3}
\text{Tr}[\Pi^{(st\otimes sgn)}(\rho\otimes H_{0})^{\otimes2}]&\propto&
%&=&3\text{Tr}[(\rho\otimes H_{0})^{\otimes2}]-\text{Tr}[(S_{(12)(34)}+...)(\rho\otimes H_{0})^{\otimes2}]\nonumber \\
%&+&\text{Tr}[(S_{(1234)}+...)(\rho\otimes H_{0})^{\otimes2}]-\text{Tr}[(S_{(12)}+...)(\rho\otimes H_{0})^{\otimes2} \nonumber \\
%&=&3\text{Tr}^{2}[H_{0}] \nonumber \\
%& &-[\sum_{ijkl}(\bra{ijkl}(\rho\otimes H_{0})^{\otimes2}\ket{jikl}+\bra{ijkl}(\rho\otimes %H_{0})^{\otimes2}\ket{klij}+\bra{ijkl}(\rho\otimes H_{0})^{\otimes2}\ket{lkji})]+...\nonumber \\
%&=&3Tr[(\rho\otimes H_{0})^{\otimes2}]+
%\sum_{abcdef}
\sum
\Big[ -\Big(\rho_{ac}\epsilon_{a}\rho_{dd}\epsilon_{f}+\rho_{ad}\epsilon_{c}\rho_{da}\epsilon_{f}+\rho_{af}\epsilon_{c}\rho_{dd}\epsilon_{a} \nonumber \\ 
& &\nonumber +\rho_{aa}\epsilon_{d}\rho_{dc}\epsilon_{f}+\rho_{aa}\epsilon_{f}\rho_{dd}\epsilon_{c}+\rho_{aa}\epsilon_{c}\rho_{df}\epsilon_{d}\Big) \nonumber \\
& &+\Big(\rho_{af}\epsilon_{a}\rho_{dc}\epsilon_{d}+\rho_{ad}\epsilon_{a}\rho_{dc}\epsilon_{d}+\rho_{af}\epsilon_{d}\rho_{da}\epsilon_{c}+\rho_{ac}\epsilon_{f}\rho_{da}\epsilon_{d}+\rho_{ad}\epsilon_{f}\rho_{da}\epsilon_{a}
+\rho_{ac} \epsilon_{d}\rho_{df}\epsilon_{a}\Big) \nonumber \\
& &-\Big(\rho_{ac}\epsilon_{a}\rho_{df}\epsilon_{d}+\rho_{ad}\epsilon_{f}\rho_{da}\epsilon_{c}+\rho_{af}\epsilon_{d}\rho_{dc}\epsilon_{a}\Big)\Big]
\end{eqnarray}

\begin{eqnarray}
%\frac{24}{2}
\text{Tr}\Big[\Pi^{(2D)}(\rho\otimes H_{0})^{\otimes2}\Big]&\propto&
%&=&2\text{Tr}\Big[(\rho\otimes H_{0})^{\otimes2}\Big]-\Big[\sum_{ijkl}(\bra{ijkl}(\rho\otimes H_{0})^{\otimes2}\ket{kijl}+\cdots \nonumber \\
%&=&2\text{Tr}\Big[(\rho\otimes H_{0})^{\otimes2}\Big]+
%\sum_{abcdef}
\sum
\Big[-\Big(\rho_{ad}\epsilon_{a}\rho_{dc}\epsilon_{f}+\rho_{ac}\epsilon_{d}\rho_{da}\epsilon_{f}+\rho_{aa}\epsilon_{f}\rho_{dc}\epsilon_{d}+\rho_{aa}\epsilon_{d}\rho_{df}\epsilon_{c} \nonumber \\ 
& &\nonumber+\rho_{af}\epsilon_{c}\rho_{da}\epsilon_{d} +\rho_{af}\epsilon_{a}\rho_{dd}\epsilon_{c}+\rho_{ac}\epsilon_{f}\rho_{dd}\epsilon_{a}\Big)+2\Big(\rho_{ac}\epsilon_{a}\rho_{df}\epsilon_{d}+\nonumber\rho_{ad}\epsilon_{f}\rho_{da}\epsilon_{c}+\rho_{af}\epsilon_{d}\rho_{dc}\epsilon_{a}\Big)\Big]
\end{eqnarray}

%------------------------------------------------------------ \\ \\
%\underline{$Tr[\Pi^{(2D)}K^{\otimes 2}\otimes K^{\dagger \otimes 2}]$'s:} \\ \\
We can now evaluate the trace over the operator $K\otimes K^\dagger$ with the projectors, $Tr[\Pi K^{\otimes 2}\otimes K^{\dagger \otimes 2}]$'s. We have the following results:
\begin{eqnarray}
\text{Tr}[\Pi^{(tr)}K^{\otimes 2}\otimes K^{\dagger \otimes 2}] %&=&\sum_{mnop}\Big(e^{i(\theta_{p}+\theta_{m}-\theta_{n}-\theta_{o})}+e^{i(\theta_{o}+\theta_{m}-\theta_{p}-\theta_{n})}+e^{i(\theta_{p}+\theta_{o}-\theta_{m}-\theta_{n})}+e^{i(\theta_{n}+\theta_{p}-\theta_{m}-\theta_{o})}\\
%\nonumber&\ &\ \ \ \ \ +e^{i(\theta_{o}+\theta_{p}-\theta_{n}-\theta_{m})}+e^{i(\theta_{n}+\theta_{o}-\theta_{p}\theta_{m})}+e^{i(\theta_{n}+\theta_{o}-\theta_{m}-\theta_{p})}+e^{i(\theta_{m}+\theta_{p}-\theta_{n}-\theta_{o})}\\
%\nonumber&\ &\ \ \ \ \ +e^{i(\theta_{m}+\theta_{o}-\theta_{p}-\theta_{n})}+e^{i(\theta_{p}+\theta_{n}-\theta_{m}-\theta_{o})} +e^{i(\theta_{o}+\theta_{n}-\theta_{p}-\theta_{m})}+e^{i(\theta_{p}+\theta_{m}-\theta_{o}-\theta_{n})}\\ \nonumber&\ &\ \ \ \ \ +e^{i(\theta_{n}+\theta_{p}-\theta_{o}-\theta_{m})}+e^{i(\theta_{m}+\theta_{n}-\theta_{p}-\theta_{p})}+e^{i(\theta_{n}+\theta_{m}-\theta_{o}-\theta_{p})}+e^{i(\theta_{o}+\theta_{n}-\theta_{m}-\theta_{p})}\\ \nonumber&\ &\ \ \ \ \ +e^{i(\theta_{p}+\theta_{n}-\theta_{o}-\theta_{m})}+e^{i(\theta_{m}+\theta_{o}-\theta_{n}-\theta_{p})}+e^{i(\theta_{m}+\theta_{p}-\theta_{o}-\theta_{n})}+e^{i(\theta_{m}+\theta_{n}-\theta_{p}-\theta_{o})}
%\\ 
%\nonumber&\ &\ \ \ \ +e^{i(\theta_{n}+\theta_{m}-\theta_{p}-\theta_{o})}+e^{i(\theta_{o}+\theta_{p}-\theta_{m}-\theta_{n})}+e^{i(\theta_{p}+\theta_{o}-\theta_{n}-\theta_{m})}+e^{i(\theta_{o}+\theta_{m}-\theta_{n}-\theta_{p})}\Big)
%\\ 
&\propto&\sum_{mnop}4 e^{i(\theta_{m}+\theta_{p}-\theta_{n}-\theta_{o})}+e^{i(\theta_{m}+\theta_{o}-\theta_{m}-\theta_{n})}+e^{i(\theta_{o}+\theta_{p}-\theta_{m}-\theta_{n})}+4 e^{i(\theta_{p}+\theta_{n}-\theta_{m}-\theta_{o})}\\
\nonumber& &\ \ \ \ \ \ +e^{i(\theta_{n}+\theta_{o}-\theta_{p}-\theta_{n})}+4 e^{i(\theta_{m}+\theta_{n}-\theta_{o}-\theta_{p})}
\end{eqnarray}

\begin{eqnarray}
\text{Tr}[\Pi^{(sig)}K^{\otimes 2}\otimes K^{\dagger \otimes 2}]&=&0
%\sum_{mnop} e^{i(\theta_{m}+\theta_{n}-\theta_{o}-\theta_{p})}+e^{i(\theta_{o}+\theta_{m}-\theta_{n}-\theta_{p})}\\ \nonumber&+&e^{i(\theta_{n}+\theta_{o}-\theta_{m}-\theta_{p})}+e^{i(\theta_{m}+\theta_{p}-\theta_{n}-\theta_{o})}+e^{i(\theta_{m}+\theta_{o}-\theta_{p}-\theta_{n})}\\ \nonumber&+&e^{i(\theta_{p}+\theta_{n}-\theta_{m}-\theta_{o})}+e^{i(\theta_{o}+\theta_{n}-\theta_{p}-\theta_{m})}+e^{i(\theta_{p}+\theta_{m}-\theta_{o}-\theta_{n})}\\ \nonumber&+&e^{i(\theta_{n}+\theta_{p}-\theta_{o}-\theta_{m})}+(e^{i(\theta_{n}+\theta_{m}-\theta_{p}-\theta_{o})}+e^{i(\theta_{o}+\theta_{p}-\theta_{m}-\theta_{n})}\\ \nonumber&+&e^{i(\theta_{p}+\theta_{o}-\theta_{n}-\theta_{m})}-[(e^{i(\theta_{p}+\theta_{m}-\theta_{n}-\theta_{o})}+e^{i(\theta_{o}+\theta_{m}-\theta_{p}-\theta_{n})}\\ \nonumber&+&e^{i(\theta_{p}+\theta_{o}-\theta_{m}-\theta_{n})}+e^{i(\theta_{n}+\theta_{p}-\theta_{m}-\theta_{o})}+e^{i(\theta_{o}+\theta_{p}-\theta_{n}-\theta_{m})}\\ \nonumber&+&e^{i(\theta_{n}+\theta_{o}-\theta_{p}\theta_{m})})+(e^{i(\theta_{n}+\theta_{m}-\theta_{o}-\theta_{p})}+e^{i(\theta_{o}+\theta_{n}-\theta_{m}-\theta_{p})}\\ \nonumber&+&e^{i(\theta_{p}+\theta_{n}-\theta_{o}-\theta_{m})}+e^{i(\theta_{m}+\theta_{o}-\theta_{n}-\theta_{p})}+e^{i(\theta_{m}+\theta_{p}-\theta_{o}-\theta_{n})}\\ \nonumber&+&e^{i(\theta_{m}+\theta_{n}-\theta_{p}-\theta_{o})})]=0
\end{eqnarray}
\begin{eqnarray}
%\frac{24}{3}
Tr[\Pi^{(st)}K^{\otimes 2}\otimes K^{\dagger \otimes 2}]\propto %=4
\sum_{mnop} \Big(e^{i(\theta_{m}+\theta_{n}-\theta_{o}-\theta_{p})}- e^{i(\theta_{o}+\theta_{p}-\theta_{m}-\theta_{n})}\Big)
\end{eqnarray}
\begin{equation}
    %\frac{24}{3}
    Tr[\Pi^{(st\otimes sgn)}K^{\otimes 2}\otimes K^{\dagger \otimes 2}]\propto -\sum_{mnop}  e^{i(\theta_{o}+\theta_{p}-\theta_{m}-\theta_{n})}
\end{equation}
\begin{eqnarray}
%\frac{24}{2}
Tr[\Pi^{(2D)}K^{\otimes 2}\otimes K^{\dagger \otimes 2}]%&=&\sum_{mnop}e^{i(\theta_{m}+\theta_{n}-\theta_{o}-\theta_{p})}[\langle mnop|kjil\rangle \langle ijkl|mnop\rangle 
%-\langle mnop | jkil\rangle \langle ijkl | mnop\rangle+...]\\ 
&\propto&\sum_{mnop} \Big[e^{i(\theta_{n}+\theta_{o}-\theta_{m}-\theta_{p})}
+e^{i(\theta_{n}+\theta_{m}-\theta_{p}-\theta_{o})}+2e^{i(\theta_{0}+\theta_{p}-\theta_{m}-\theta_{n})}+2e^{i(\theta_{p}+\theta_{o}-\theta_{n}-\theta_{m})} \nonumber \\
&\ &\ \ \ \ -\Big(e^{i(\theta_{m}+\theta_{p}-\theta_{n}-\theta_{o})}+e^{i(\theta_{m}+\theta_{o}-\theta_{p}-\theta_{n})}+e^{i(\theta_{o}+\theta_{n}-\theta_{p}-\theta_{m})}+e^{i(\theta_{p}+\theta_{m}-\theta_{o}-\theta_{n})} \nonumber \\
& &\ \ \ \ \ \ \ \ \ \ \ +e^{i(\theta_{n}+\theta_{p}-\theta_{o}-\theta_{m})}\Big)\Big]
\end{eqnarray}
%------------------------------------------------------------ \\ \\
We now consider the traces of the projectors alone, $Tr[\Pi]$'s. It is not hard to see that for large values of $n$, we have

\begin{eqnarray}
\text{Tr}[\Pi^{(tr)}]&\propto& n^4  \\%&=&\frac{1}{24}\sum_{ijkl}\Big[\langle ijkl | lijk\rangle+\langle ijkl | kilj \rangle + \langle ijkl | lkij \rangle \\ 
%\nonumber& &\ \ \ \ \ \ \ \ \ \ \ \ + \langle ijkl | jlik \rangle + \langle ijkl | klji \rangle + \langle ijkl | jkli  \rangle+\langle ijkl | kijl \rangle \\ 
%\nonumber& &\ \ \ \ \ \ \ \ \ \ \ \ +  \langle  ijkl | jkil \rangle + \langle ijkl | iljk \rangle + \langle ijkl | iklj \rangle +\langle ijkl | ljik \rangle + \langle ijkl | kjli \rangle \\ 
%\nonumber& &\ \ \ \ \ \ \ \ \ \ \ \ +  \langle ijkl | likj \rangle + \langle ijkl | jlki \rangle + \langle ijkl | ijkl \rangle +\langle ijkl | jikl \rangle \\ 
%\nonumber& &\ \ \ \ \ \ \ \ \ \ \ \ +   \langle ijkl | kjil \rangle +\langle ijkl | ljki \rangle +\langle ijkl | ikjl \rangle + \langle ijkl | ilkj \rangle + \langle ijkl | ijlk \rangle\\ 
%\nonumber & &\ \ \ \ \ \ \ \ \ \ \ \ +  \langle ijkl | jilk \rangle + \langle ijkl | klij \rangle + \langle ijkl | lkji \rangle\Big] \\
%&=& \frac{1}{24}[n^{8}+6n^{6}+11n^{4}+6n^{2}]=\frac{1}{24}[n^{2}(n^{2}+1)(n^{2}+2)(n^{2}+3)] \simeq \frac{n^{8}}{24}
%\end{eqnarray}
%\begin{eqnarray}
\text{Tr}[\Pi^{(sig)}]&\propto& n^4\\%&=&\frac{1}{24}\sum_{ijkl}\Big[\langle ijkl | ijkl \rangle + (\langle ijkl | kijl \rangle + \langle  ijkl | jkil \rangle \\ 
%\nonumber& &\ \ \ \ \ \ \ \ \ \ \ \ + \langle ijkl | iljk \rangle + \langle ijkl | iklj \rangle +\langle ijkl | ljik \rangle + \langle ijkl | kjli \rangle \\ 
%\nonumber& &\ \ \ \ \ \ \ \ \ \ \ \ + \langle ijkl | likj \rangle + \langle ijkl | jlki \rangle) + (\langle ijkl | jilk \rangle + \langle ijkl | klij \rangle + \langle ijkl | lkji \rangle) \\ 
%\nonumber& &\ \ \ \ \ \ \ \ \ \ \ \ -(\langle ijkl | lijk\rangle+\langle ijkl | kilj \rangle + \langle ijkl | lkij \rangle + \langle ijkl | jlik \rangle + \langle ijkl | klji \rangle \\ 
%\nonumber& &\ \ \ \ \ \ \ \ \ \ \ \ + \langle ijkl | jkli  \rangle) - (\langle ijkl | jikl \rangle +  \langle ijkl | kjil \rangle +\langle ijkl | ljki \rangle 
%\\ \nonumber& &\ \ \ \ \ \ \ \ \ \ \ \ +\langle ijkl | ikjl \rangle + \langle ijkl | ilkj \rangle + \langle ijkl | ijlk \rangle)\Big]\\ &=&\frac{1}{24}(n^{8}+8n^{4}+3n^{4}-(6n^{2}+6n^{6})=\frac{1}{24}[n^{2}(n^{2}-1)(n^{2}-2)(n^{2}-3)]\simeq \frac{n^{8}}{24}
%\end{eqnarray}
%\begin{eqnarray}
\text{Tr}[\Pi^{(st)}]&\propto& n^4 \\%\frac{3}{24}\sum_{ijkl} \Big[]3\langle ijkl | ijkl \rangle + (\langle ijkl | jikl \rangle +  \langle ijkl | kjil \rangle \\ 
%\nonumber& &\ \ \ \ \ \ \ \ \ \ \ \ +\langle ijkl | ljki \rangle +\langle ijkl | ikjl \rangle + \langle ijkl | ilkj \rangle + \langle ijkl | ijlk \rangle) \\ \nonumber& &\ \ \ \ \ \ \ \ \ \ \ \ - (\langle ijkl | lijk\rangle+\langle ijkl | kilj \rangle + \langle ijkl | lkij \rangle + \langle ijkl | jlik \rangle + \langle ijkl | klji \rangle \\ 
%\nonumber& &\ \ \ \ \ \ \ \ \ \ \ \ + \langle ijkl | jkli  \rangle) - (\langle ijkl | jilk \rangle + \langle ijkl | klij \rangle + \langle ijkl | lkji \rangle)\Big]\\
%&=&\frac{3}{24} [3n^{8}+6n^{6}-6n^{2}-3n^{4}]\approx \frac{3}{24 } n^4
%\end{eqnarray}
%\begin{eqnarray}
\text{Tr}[\Pi^{(st\otimes sgn)}]&\propto&n^4\\
%&=& \frac{3}{24}\sum_{ijkl}\Big[3\langle ijkl | ijkl \rangle - (\langle ijkl | jikl \rangle +  \langle ijkl | kjil \rangle \\ 
%\nonumber & &\ \ \ \ \ \ \ \ \ \ \ \ +\langle ijkl | ljki \rangle +\langle ijkl | ikjl \rangle + \langle ijkl | ilkj \rangle + \langle ijkl | ijlk \rangle) \\
%\nonumber& &\ \ \ \ \ \ \ \ \ \ \ \ + (\langle ijkl | lijk\rangle+\langle ijkl | kilj \rangle + \langle ijkl | lkij \rangle + \langle ijkl | jlik \rangle + \langle ijkl | klji \rangle \\ 
%\nonumber & &\ \ \ \ \ \ \ \ \ \ \ \ + \langle ijkl | jkli  \rangle) - (\langle ijkl | jilk \rangle + \langle ijkl | klij \rangle + \langle ijkl | lkji \rangle)\Big]\\
%&=&\frac{3}{24}[3n^{8}-6n^{6}+6n^{2}-3n^{4}]\approx \frac{3}{24} n^4
%\end{eqnarray}
%\begin{eqnarray}
\text{Tr}[\Pi^{(2D)}]&\propto& n^4%&=& \frac{2}{24}\sum_{ijkl} \Big[2\langle ijkl | ijkl \rangle - \langle ijkl | jikl \rangle +  \langle ijkl | kjil \rangle \\ 
%\nonumber& &\ \ \ \ \ \ \ \ \ \ \ \ +\langle ijkl | ljki \rangle +\langle ijkl | ikjl \rangle + \langle ijkl | ilkj \rangle + \langle ijkl | ijlk \rangle \\
%\nonumber& &\ \ \ \ \ \ \ \ \ \ \ \ + (\langle ijkl | ijkl \rangle) +2(\langle ijkl | jilk \rangle + \langle ijkl | klij %\rangle + \langle ijkl | lkji \rangle)\Big]\\
%&=&\frac{2}{24}(2n^{8}-8n^{4}+6n^{4}) =\frac{2}{24}[2n^{8}-2n^{4}]\approx \frac{2}{24}n^4
\end{eqnarray}
%------------------------------------------------------------ \\ \\
At this point we can calculate the
 average fluctuations, which can be written as 
\begin{equation}
    F=F_{\Pi^{(tr)}}+F_{\Pi^{(sig)}}+F_{\Pi^{(st)}}+F_{\Pi^{(st\otimes sgn)}}+F_{\Pi^{(2D)}}
\end{equation}
\begin{eqnarray}
F_{\Pi^{(tr)}}&\propto & \frac{1}{n^4} %\Big[\frac{1}{24}[d^{2}(d^{2}+1)(d^{2}+2)(d^{2}+3)]\Big]^{-1}
\Big[\sum_{mnop}4e^{i(\theta_{m}+\theta_{p}-\theta_{n}-\theta_{o})}+4e^{i(\theta_{m}+\theta_{o}-\theta_{m}-\theta_{n})}+4 e^{i(\theta_{o}+\theta_{p}-\theta_{m}-\theta_{n})} \\
& &\ \ \ \ \ \ \ \ \ \ \ \ \ \ \ \ \ \ \ \ \ \ \ \ \ \ \ \ \ \ \ \ \ \ \ \ \ \ \ \ \ \ \ \ \ \ \ \ \ \ \ \ \ \ \ \ \ \ \ \ \ \ \ +4e^{i(\theta_{p}+\theta_{n}-\theta_{m}-\theta_{o})}+4e^{i(\theta_{n}+\theta_{o}-\theta_{p}-\theta_{n})}+4e^{i(\theta_{m}+\theta_{n}-\theta_{o}-\theta_{p})}\Big]\\ \nonumber& &\ \ \ \ \ \ \ \ \ \cdot
%\sum_{abcdef}
\sum
%\Big[ \rho_{aa}\epsilon_{c}\rho_{dd}\epsilon_{f} + \rho_{ad}\epsilon_{a}\rho_{de}\epsilon_{f}+\rho_{ac}\epsilon_{d}\rho_{da}\epsilon_{f}
%+\rho_{aa}\epsilon_{f}\rho_{dc}\epsilon_{d}+\rho_{aa}\epsilon_{d}\rho_{df}\epsilon_{c}+\rho_{af}\epsilon_{c}\rho_{da}\epsilon_{d}+\rho_{ad}\epsilon_{c}\rho_{df}\epsilon_{a} \\ 
%\nonumber& &\ \ \ \ \ \ \ \ \ \ +\rho_{af}\epsilon_{a}\rho_{dd}\epsilon_{c}+\rho_{ac}\epsilon_{f}\rho_{dd}\epsilon_{a}+\rho_{ac}\epsilon_{a}\rho_{df}\epsilon_{d}+\rho_{ad}\epsilon_{f}\rho_{da}\epsilon_{c}+%\rho_{af}\epsilon_{d}\rho_{dc}\epsilon_{a}-[(\rho_{af}\epsilon_{a}\rho_{dc}\epsilon_{d}+\rho_{ad}\epsilon_{a}\rho_{df}\epsilon_{c}+\rho_{af}\epsilon_{d}\rho_{da}\epsilon_{c}\\ 
%\nonumber& &\ \ \ \ \ \ \ \ \ \ +\rho_{ac}\epsilon_{f}\rho_{da}\epsilon_{d}+\rho_{ad}\epsilon_{f}\rho_{da}\epsilon_{a}+\rho_{ac}\epsilon_{d}\rho_{df}\epsilon_{a})+(\rho_{ac}\epsilon_{a}\rho_{dd}\epsilon_{f}+%\rho_{ad}\epsilon_{e}\rho_{da}\epsilon_{f}+\rho_{af}\epsilon_{c}\rho_{dd}\epsilon_{a}
%\\ \nonumber& &\ \ \ \ \ \ \ \ \ \ +\rho_{aa}\epsilon_{d}\rho_{dc}\epsilon_{f}+\rho_{aa}\epsilon_{f}\rho_{dd}\epsilon_{c}+\rho_{aa}\epsilon_{c}\rho_{df}\epsilon_{d})]\Big] \\
\Big[ \rho_{aa}\epsilon_{c}\rho_{dd}\epsilon_{f} + \rho_{ad}\epsilon_{a}\rho_{de}\epsilon_{f}+\rho_{ac}\epsilon_{d}\rho_{da}\epsilon_{f}
+\rho_{aa}\epsilon_{f}\rho_{dc}\epsilon_{d}+\rho_{aa}\epsilon_{d}\rho_{df}\epsilon_{c}+\rho_{af}\epsilon_{c}\rho_{da}\epsilon_{d}+\rho_{ad}\epsilon_{c}\rho_{df}\epsilon_{a} \\ 
\nonumber& &\ \ \ \ \ \ \ \ \ \ +\rho_{af}\epsilon_{a}\rho_{dd}\epsilon_{c}+\rho_{ac}\epsilon_{f}\rho_{dd}\epsilon_{a}+\rho_{ac}\epsilon_{a}\rho_{df}\epsilon_{d}+\rho_{ad}\epsilon_{f}\rho_{da}\epsilon_{c}+\rho_{af}\epsilon_{d}\rho_{dc}\epsilon_{a}+\rho_{af}\epsilon_{a}\rho_{dc}\epsilon_{d}+\rho_{ad}\epsilon_{a}\rho_{df}\epsilon_{c}\\ 
\nonumber& &\ \ \ \ \ \ \ \ \ \ +\rho_{af}\epsilon_{d}\rho_{da}\epsilon_{c}+\rho_{ac}\epsilon_{f}\rho_{da}\epsilon_{d}+\rho_{ad}\epsilon_{f}\rho_{da}\epsilon_{a}+\rho_{ac}\epsilon_{d}\rho_{df}\epsilon_{a}+\rho_{ac}\epsilon_{a}\rho_{dd}\epsilon_{f}+\rho_{ad}\epsilon_{e}\rho_{da}\epsilon_{f}+\rho_{af}\epsilon_{c}\rho_{dd}\epsilon_{a}
\\ \nonumber& &\ \ \ \ \ \ \ \ \ \ +\rho_{aa}\epsilon_{d}\rho_{dc}\epsilon_{f}+\rho_{aa}\epsilon_{f}\rho_{dd}\epsilon_{c}+\rho_{aa}\epsilon_{c}\rho_{df}\epsilon_{d}\Big] 
\end{eqnarray}
\ba
F_{\Pi^{(sig)}}=0
\ea
\begin{eqnarray}
F_{\Pi^{(st)}}&\propto& \frac{1}{n^4}
%\Big[\frac{3}{24} [3d^{8}+6d^{6}-6d^{2}-3d^{4}]\Big]^{-1}
%[\frac{3}{24}
[\sum_{mnop}4e^{i(\theta_{m}+\theta_{n}-\theta_{o}-\theta_{p})}-4e^{i(\theta_{o}+\theta_{p}-\theta_{m}-\theta_{n})}]\\ 
& &\nonumber\Big[3\text{Tr}[(\rho\otimes H_{0})^{\otimes2}]-
%\sum_{abcdef}
\sum
\Big( (\rho_{ac}\epsilon_{a}\rho_{dd}\epsilon_{f}+\rho_{ad}\epsilon_{c}\rho_{da}\epsilon_{f}+\rho_{af}\epsilon_{c}\rho_{dd}\epsilon_{a}+ \rho_{aa}\epsilon_{d}\rho_{dc}\epsilon_{f}+\rho_{aa}\epsilon_{f}\rho_{dd}\epsilon_{c}+\rho_{aa}\epsilon_{c}\rho_{df}\epsilon_{d})\\ \nonumber& &\ \ \ \ \ \ \ \ \ \ -(\rho_{af}\epsilon_{a}\rho_{dc}\epsilon_{d}+\rho_{ad}\epsilon_{a}\rho_{dc}\epsilon_{d}+\rho_{af}\epsilon_{d}\rho_{da}\epsilon_{c}+\rho_{ac}\epsilon_{f}\rho_{da}\epsilon_{d}+\rho_{ad}\epsilon_{f}\rho_{da}\epsilon_{a}+\rho_{ac}\epsilon_{d}\rho_{df}\epsilon_{a})\\ 
\nonumber& &\ \ \ \ \ \ \ \ \ \ -(\rho_{ac}\epsilon_{a}\rho_{df}\epsilon_{d}+\nonumber\rho_{ad}\epsilon_{f}\rho_{da}\epsilon_{c}+\rho_{af}\epsilon_{d}\rho_{dc}\epsilon_{a})\Big)\Big]
\end{eqnarray}
\begin{eqnarray}
F_{\Pi^{(st\otimes sgn)}}
&\propto& \frac{1}{n^4}%\Big[\frac{3}{24}[3d^{8}-6d^{6}+6d^{2}-3d^{4}]\Big]^{-1}
\Big[
%-\frac{3}{24}
\sum_{mnop}  e^{i(\theta_{o}+\theta_{p}-\theta_{m}-\theta_{n})}\Big]\\ 
& &\nonumber \cdot \frac{3}{24}\Big[3Tr[(\rho\otimes H_{0})^{\otimes2}]+[
%\sum_{abcdef} 
\sum 
-(\rho_{ac}\epsilon_{a}\rho_{dd}\epsilon_{f}+\rho_{ad}\epsilon_{c}\rho_{da}\epsilon_{f}+\rho_{af}\epsilon_{c}\rho_{dd}\epsilon_{a}+\\ 
& &\nonumber \rho_{aa}\epsilon_{d}\rho_{dc}\epsilon_{f}+\rho_{aa}\epsilon_{f}\rho_{dd}\epsilon_{c}+\rho_{aa}\epsilon_{c}\rho_{df}\epsilon_{d})+(\rho_{af}\epsilon_{a}\rho_{dc}\epsilon_{d}+\rho_{ad}\epsilon_{a}\rho_{dc}\epsilon_{d}+\rho_{af}\epsilon_{d}\rho_{da}\epsilon_{c}+\rho_{ac}\epsilon_{f}\rho_{da}\epsilon_{d}+\\
& &\nonumber\rho_{ad}\epsilon_{f}\rho_{da}\epsilon_{a}+\rho_{ac}\epsilon_{d}\rho_{df}\epsilon_{a})-(\rho_{ac}\epsilon_{a}\rho_{df}\epsilon_{d}+\rho_{ad}\epsilon_{f}\rho_{da}\epsilon_{c}+\rho_{af}\epsilon_{d}\rho_{dc}\epsilon_{a})]\Big]
\end{eqnarray}
\begin{eqnarray}
F_{\Pi^{(2D)}}&\propto&\frac{1}{n^4}
%\Big[\frac{2}{24}[2d^{8}-2d^{4}]\Big]^{-1}[\frac{2}{24}
\sum_{mnop} \Big(
e^{i(\theta_{n}+\theta_{o}-\theta_{m}-\theta_{p})}-(e^{i(\theta_{m}+\theta_{p}-\theta_{n}-\theta_{o})}\\ \nonumber\ \ \ \ \ \ \ \ \ &+&e^{i(\theta_{m}+\theta_{o}-\theta_{p}-\theta_{n})}+e^{i(\theta_{o}+\theta_{n}-\theta_{p}-\theta_{m})}+e^{i(\theta_{p}+\theta_{m}-\theta_{o}-\theta_{n})}+e^{i(\theta_{n}+\theta_{p}-\theta_{o}-\theta_{m})})\\ 
\nonumber\ \ \ \ \ \ \ \ \ &+&2e^{i(\theta_{n}+\theta_{m}-\theta_{p}-\theta_{o})}+2e^{i(\theta_{0}+\theta_{p}-\theta_{m}-\theta_{n})}+2e^{i(\theta_{p}+\theta_{o}-\theta_{n}-\theta_{m})}\Big)\\ \nonumber
&\cdot &\Big[2\text{Tr}[(\rho\otimes H_{0})^{\otimes2}]+
%\sum_{abcdef}
\sum
-(\rho_{ad}\epsilon_{a}\rho_{dc}\epsilon_{f}+\rho_{ac}\epsilon_{d}\rho_{da}\epsilon_{f}+\rho_{aa}\epsilon_{f}\rho_{dc}\epsilon_{d}+\rho_{aa}\epsilon_{d}\rho_{df}\epsilon_{c}+\rho_{af}\epsilon_{c}\rho_{da}\epsilon_{d}+\\
& &\ \ \ \ \ \ \ \ \ \  \nonumber \rho_{af}\epsilon_{a}\rho_{dd}\epsilon_{c}+\rho_{ac}\epsilon_{f}\rho_{dd}\epsilon_{a})+2(\rho_{ac}\epsilon_{a}\rho_{df}\epsilon_{d}+\nonumber\rho_{ad}\epsilon_{f}\rho_{da}\epsilon_{c}+\rho_{af}\epsilon_{d}\rho_{dc}\epsilon_{a})\Big]
\end{eqnarray}

\subsection{Concentration bound}\label{appendix8b}
Let us now consider an upper bound for the  non-zero fluctuation terms based on general grounds and on the von Neumann inequality \cite{perspectrive}. Let $A$ and $B$ be hermitean matrices with eigenvalues values of $a_i \geq a_{i-1}$'s and $b_{i}\geq b_{i-1}$. Then, we have
\begin{equation}
    |\text{Tr}(AB)|\leq \sum_{i=1}^n a_i b_i.
\end{equation}
Let us now assume that $A$ is a projector with $k$ non-zero eigenvalues. Then the inequality implies that
\begin{equation}
|\text{Tr}(\Pi B)|\leq \sum_{j=n-k}^n b_{j}
\end{equation}
where $b_{n}\cdots b_{n-k}$ are the highest $k$'s eigenvalues values of $B$.
We thus need to focus on the singular values of $(\rho\otimes H)^{\otimes 2}$. 
The eigenvalues of $\rho \otimes H$, are $e_{ij}=p_i \epsilon_{j}$, and the eigenvalues $(\rho \otimes H)^{\otimes 2}$ are $e_{ijkl}=p_i \epsilon_{j} p_k \epsilon_{l}$. Since $p_i\leq 1$ in the most general case, $e_{ijkl}$ is upper-bounded by $\epsilon_{max}^2$. 
We thus have that a conservative upper bound is given by 
\begin{equation}
|\text{Tr}\Big(\Pi (\rho \otimes H)^{\otimes 2}\Big)|\leq k^2 \epsilon_{max}^2
\end{equation}
where $k$ is the dimension of the non-zero subspace of the projector operator.
Since each term of the trace is divided by the dimension of the projector operator and we have four non-zero terms, we have
\begin{equation}
F\leq 4 \epsilon_{max}^2
\end{equation}
in the most general case. However, the bound $p_i\leq 1$ is very loose. If the $p_i\leq \frac{\gamma}{n}$, we have 
\begin{equation}
F\leq 4 \gamma^2 \frac{\epsilon_{max}^2}{n^2}
\end{equation}
and thus there is concentration. For instance, we have concentration if we have that $\rho$ is a mixed sate. A stronger bound can be done by using the expressions we derived. We see from the bound above that this is not enough to prove concentration. However, the concentration can be proven if the take advantage of the structure of the fluctuations in terms of the density matrix.

We now provide an alternative proof of the same statement. From the previous subsection, we see that the work fluctuations $F$ can be upper bounded as 
\ba
F\le C n^{-4} M(n) kn^2 
\ea
where $C$ is a $O(1)$ constant counting the number of all the terms in $F$, $M(n)$ is an upper  bound over al the terms of the type $\left|\sum_{mnop}  e^{i(\theta_{o}+\theta_{p}-\theta_{m}-\theta_{n})}\right|$ and $ kn^2 $ is the upper bound to the terms containing the $\rho,H_0$:
\ba
|\tr [ \Pi_x (\rho\otimes H_0)^{\otimes 2}]|\le |\tr [(\rho\otimes H_0)^{\otimes 2}]| = (\tr \rho)^2(\tr H_0)^2 = (\tr H_0)^2\le k n^2
\ea
which is true because the projector operators are positive. Putting things together, we obtain 
\ba
F \le C' M(n) n^{-2}
\ea
with a new constant $C'$. The fluctuations are thus ruled by $M(n)$. One can design quantum batteries with large fluctuations. However, on average these fluctuations go to zero. Indeed, it should not be surprising that the sum over (the sum of) random phases goes to zero. For random unitaries we need to use the ensemble of circulant unitary matrices (CUE). Numerical evaluation  (see in Fig. \ref{fig:concentration}) shows that $M(n)$ is concentrated around zero for large dimension $n$.
\begin{figure}
    \centering
    \includegraphics[scale=0.4]{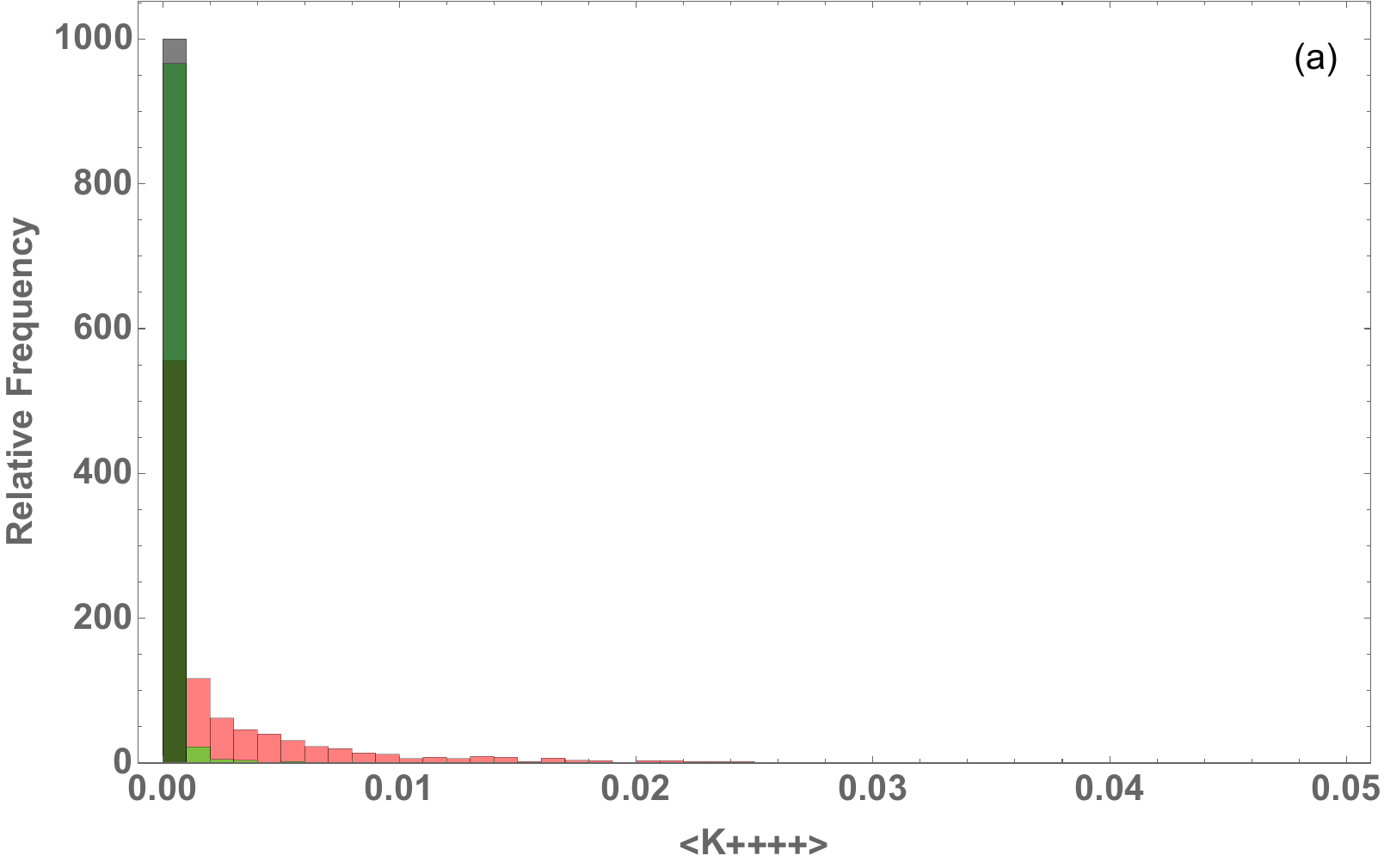}\\
    \includegraphics[scale=0.4]{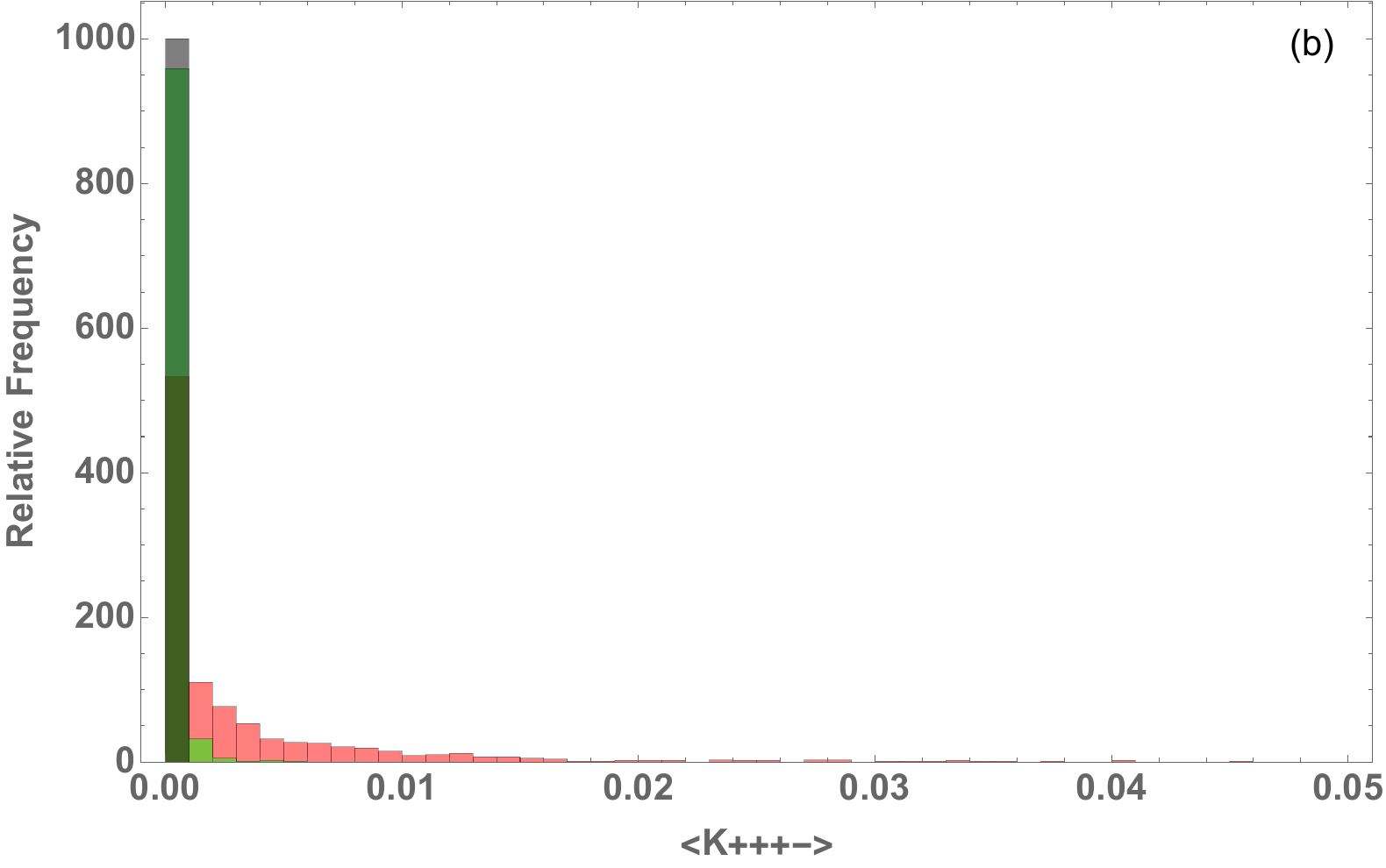}\\
    \includegraphics[scale=0.4]{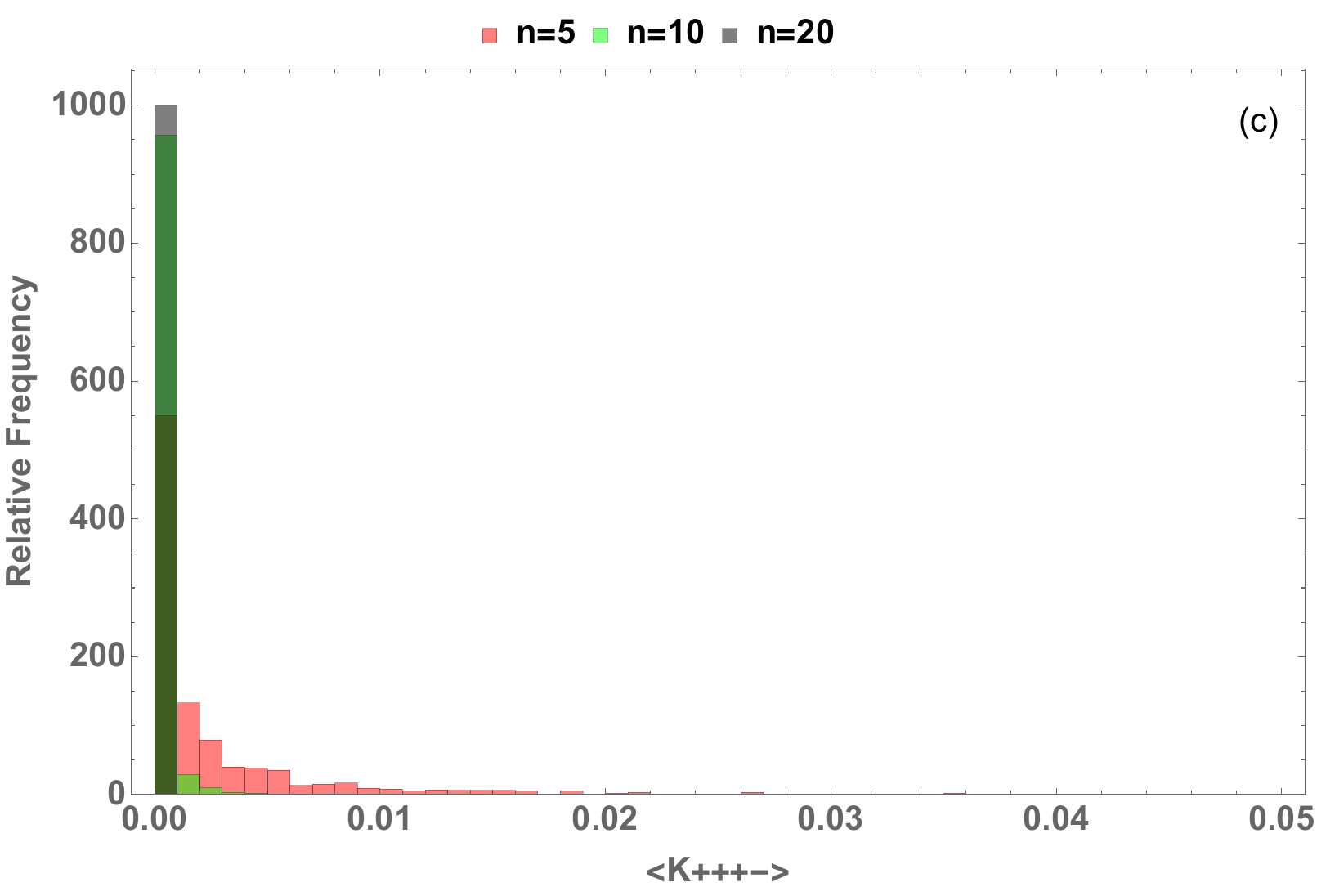}
    \caption{Frequency distribution of the average $\langle e^{\sum_{j=1}^4 \pm i\theta_j}\rangle$ for circulant unitary matrices (CUE) over $M=1000$ samples. We see that the distribution is strongly peaked around the value of $\langle K^2\rangle=0$, for the three terms with 3 possible signature in the exponent for (a),(b),(c), which is what is necessary for the proof of our concentration at least in the case of CUE. The last peak is just a binning artifact.}
    \label{fig:concentration}
\end{figure}
%to sum all it up, there is one order one constant $g$ such that
%\ba
%|F| \le \frac{g}{n^2}
%\ea
%A more elegant way to obtain the bound is via:
%\ba
%|\tr [ \Pi_x (\rho\otimes H_0)^{\otimes 2}]|\le |\tr [(\rho\otimes H_0)^{\otimes 2}]| = (\tr \rho)^2(\tr H_0)^2 = (\tr H_0)^2\le k n^2.
%\ea
%The result above proves the concentration for arbitrary density matrices $\rho$.

%%%%%%%%%%%%%%%%
\subsection{Jaynes-Cummings model}\label{appendix8c}
As seen in eqn. (\ref{avwork}), the average work depends only on the value of the eigenvalues of the Unitary evolution operator $K$.
Let us consider the case of an optical cavity interacting with a 2-state system.
 The optical cavity with the two state system (an atom)  $\mbox{span}(|g\rangle, |e\rangle)$ can be described within the rotating-wave approximation using the Jaynes-Cumming Hamiltonian:
%%%%%%%%%%%%%%%%
\begin{equation}
    H=\omega a^\dagger a+ \frac{\Omega}{2} \sigma_z+g (t) (a\sigma_++ a^\dagger \sigma_-) \equiv H_0 + V(t) 
\end{equation}
It is immediate to see that $[H,a^+ a+\sigma_z]=0$. Specifically, we focus on the interaction picture, in which $H_I=R H R^\dagger$, where (in the rotating frame) we have $R=e^{- i\omega t (a^\dagger a+\frac{\sigma_z}{2})}$, and one has a Hamiltonian described by $H_I= R H R^\dagger$, with 
\begin{equation}
    H_I=\frac{\Omega-\omega}{2} \sigma_z+ g(t)(a^\dagger \sigma_-+ \sigma_+ a)
\end{equation}
We define $\Delta=\Omega-\omega$.
The operators $a$ and $a^\dagger$ act on the electromagnetic field, while $\sigma$'s act on the two-level system.
We have
\begin{equation}
    \sigma_+=|e\rangle\langle g|,\ \ \ \ \  \sigma_-=|g\rangle\langle e|.
\end{equation}
%%%%%%
We now consider a wave function of the form
\begin{equation}
    |\psi(t)\rangle=\sum_{n=0}^R C_n(t) |n\rangle \otimes |e\rangle+ D_n(t) |n+1\rangle \otimes |g\rangle,
\end{equation}
where we will send $R\rightarrow \infty$ at the end of the calculation.
The time evolution of this system is given by the Schroedinger equation (in the interaction picture), which is of the form:
\begin{equation}
    i \partial_t |\psi(t)\rangle= H_I |\psi(t)\rangle
\end{equation}
Which is not hard to see that it can be written as
\begin{eqnarray}
    i \partial_t \left(\begin{array}{c} C_n(t) \\ D_n(t) \end{array} \right)&=& V  \left(\begin{array}{c} C_n(t) \\ D_n(t) \end{array} \right) \nonumber \\
    &=&\left(\begin{array}{cc} \frac{\Delta}{2} & \sqrt{n+1} g \\ \sqrt{n+1} g & -\frac{\Delta}{2}  \end{array} \right) \left(\begin{array}{c} C_n(t) \\ D_n(t) \end{array} \right)
\end{eqnarray}
whose solution is given by
\begin{equation}
    |\psi_n(t)\rangle= \mathcal T e^{-i \int^t V(t^\prime) dt^\prime } |\psi_n(0)\rangle
\end{equation}
We note that $V(t^\prime) V(t)\neq V(t) V(t^\prime)$ in the case of a time dependent interaction Hamiltonian. In fact, we see that on the $n-$th subspace of the wave function, given the definition $W(\Delta,g)=\Delta g^\prime- g\Delta ^\prime$ of the wronskian of the functions $\Delta$ and $g$, we have
\begin{equation}
    [V_r(t^\prime),V_r(t)]=\left(\begin{array}{cc} 0 &  W(\Delta,g)\sqrt{r+1}\\ W(\Delta,g)\sqrt{r+1} & 0 \end{array} \right)
\end{equation}
from which we observe that we can have a time dependent and commuting (at all times) Hamiltonian if we have the condition
\begin{equation}
    \Delta g^\prime= g \Delta^\prime.
\end{equation}
which can be satisfied if
\begin{equation}
    \frac{\Delta(t)}{\Delta(t^\prime)}=\frac{g(t)}{g(t^\prime)}=e^{M (t-t^\prime)}
\end{equation}
for a constant $M$. In this case, the time ordering can be removed and we can write
\begin{equation}
    \int^t_{t_0} V_r(t^\prime) dt^\prime= \left(\begin{array}{cc} \frac{\Delta}{2} & \sqrt{r+1} g \\ \sqrt{r+1} g & -\frac{\Delta}{2}  \end{array} \right) \frac{e^{M t}-e^{M t_0}}{M}
\end{equation}
The Stone operator in this case can also be written explicitly on each subspace. 
It can be shown that in each $r$-th subspaces
\begin{eqnarray}
    e^{- i \int^t_{t_0} V_r(t^\prime) dt^\prime}&=& \mathbb{I} \sum_{k=0} ^\infty \frac{(-1)^k \beta_r(t) ^{2k}}{(2k)!}  \nonumber \\
    &-& i\frac{\hat \sigma_x}{\beta_r(t)} \sum_{k=0}^\infty (-1)^k \frac{\beta_r(t)^{2k+1}}{(2k+1)!}
\end{eqnarray}
and where
\begin{equation}
    \beta_r(t)=\left(\frac{\Delta_0^2}{4}+ g^2 (r+1) \right)(\frac{e^{M t}-e^{M t_0}}{M})^2.
\end{equation}
Thus, the Stone operator which describes the time evolution on the $r$th subspace is given by
%\begin{widetext}
\begin{equation}
    e^{- i \int^t_{t_0} V_r(t^\prime) dt^\prime}=\left(\begin{array}{cc}
         \cos(\beta_r(t))-\frac{i\Delta}{2 \beta_r(t)}\sin(\beta_r(t)) &   -i \frac{g \sqrt{r+1}}{\beta_r(t)}\sin(\beta_r(t)) \\
        -i \frac{g \sqrt{r+1}}{\beta_r(t)}\sin(\beta_r(t)) & \cos(\beta_r(t))+\frac{i\Delta}{2 \beta_r(t)}\sin(\beta_r(t))
    \end{array} \right)
\end{equation}
%\end{widetext}
We now focus on the eigenvalues of the matrix above, which must be of the form $e^{i\theta_k}$. For a matrix of the type
\begin{equation}
    \left(\begin{array}{cc}
         a-i d &   -i c\\
        -i c & a+i d
    \end{array} \right),
\end{equation}
the eigenvalues are known exactly and are of the form $\lambda_{\pm}=a\pm i \sqrt{c^2+d^2}$. It is immediate to see that the eigenvalues are complex, and have norm $1$. The phases are given by $\pm \theta_{k}\equiv \pm \beta_k(t)$. We thus find that
\begin{equation}
    \theta_k-\theta_m=g_0^2 (k-m) \left(\frac{e^{M t}-e^{M t_0}}{M^2} \right)^2
\end{equation}
which is what we need for the evaluation for the work in the main text.
We can now plug this result into eqn. (\ref{avwork}), which reads
\ba
  \langle W(t)\rangle_V &=&E_0\left(1-\frac{2 \sum_{j\neq k} \cos(\theta_j-\theta_k)+ 1-n}{n^2-1}\right)  \nonumber \\
    &-& \frac{n^2-(2 \sum_{j\neq k} \cos(\theta_j-\theta_k)+ n)}{n^2-1} \frac{\tr(H_0)}{n}\nonumber 
\ea
where here $n=2 R$, and $R$ is the number of modes of the electric field.
Let us call $\alpha=g_0^2 \left(\frac{e^{M t}-e^{M t_0}}{M^2} \right)^2$.
We thus need to calculate $\sum_{j\neq k} \cos(\alpha(j-k))$. 
Thankfully, this sum is known, and is given by
%\begin{widetext}
\begin{eqnarray}\label{thankfully}
    \tilde Q(\alpha)\equiv \sum_{i=1}^n \sum_{j=i+1}^n \cos\left(\alpha (i-j)\right)&=&\frac{1}{4} \big(\cos \left(\frac{\alpha -\pi }{2}\right) \csc \left(\frac{\alpha }{2}\right)-\sin \left(\frac{\alpha -\pi }{2}\right) \cot \left(\frac{\alpha }{2}\right) \csc \left(\frac{\alpha }{2}\right) \nonumber \\
    &-&
    \csc \left(\frac{\alpha
   }{2}\right) \cos \left(\frac{1}{2} (\alpha -2 \alpha  n-\pi )\right) 
   +\cot \left(\frac{\alpha }{2}\right) \csc \left(\frac{\alpha }{2}\right) \sin \left(\frac{1}{2} (\alpha -2 \alpha  n-\pi )\right)-2 n\big),\nonumber \\
\end{eqnarray}
%\end{widetext}
from which we obtain:
\begin{eqnarray}
\langle W(t)\rangle_V &=&E_0\left(1-\frac{4 \tilde Q(\alpha_t)+ 1-n}{n^2-1}\right)  \nonumber \\
    &-& \frac{n^2-(4 \tilde Q(\alpha_t)+ n)}{n^2-1} \frac{\tr(H_0)}{n}\nonumber  \\
    &=& W_0-  \tilde Q(\alpha_t) W_1
\end{eqnarray}
with
\begin{eqnarray}
    W_0&=&E_0\left(1+\frac{ 1}{n+1}\right)- \frac{\tr(H_0)}{n+1} \nonumber \\
    W_1&=& \frac{4 }{n^2-1}( \frac{\tr(H_0)}{n}-E_0)
\end{eqnarray}
We thus see that the time dependence of the work enters only in $\tilde Q\left(\alpha(t)\right)$.

In order to calculate the times at which the revivals occur, we write $\tilde Q(\alpha)$ in terms of $\alpha(t)=2\pi z(t)$. We thus have a simpler formula:
\begin{equation}
\tilde Q\left(r(t)\right)=\frac{1}{4} \csc ^2(\pi  z(t)) (n \cos (2 \pi  z(t))-\cos (2 \pi  n z(t))-n+1).
\end{equation}
It is not hard to see that revivals occur for $z_k=k$ with $k\in\mathbb{N}$, thus for $\alpha_k$ a multiple of $2\pi$. We now have that 
\begin{equation}
\alpha_k=g_0^2 \left(\frac{e^{M t_k}-e^{M t_0}}{M^2} \right)^2=2\pi k
\end{equation}
for $k\in\mathbb{N}$, from which we get the revival times
\begin{equation}
t_k=\frac{\log \left(\frac{g_0 e^{t_0 M }+\sqrt{2 \pi } M \sqrt{k}  }{g_0}\right)}{M
   }
\end{equation}
as a function of $M$ and $g_0$.

%%%
%%%%%%%%%%%%%%%%%%%%%%%%%%%%%%
%FIGURE
%%%%%%%%%%%%%%%%%%%%%%%%%%%%%%

%%%%%%%%%%%%%%%%%%%%%%%%%%%%%%

%%%%%%%%%%%%%%%%%%%%%%%%%%%%%%
\subsection{Time dependent perturbation theory}\label{appendix8d}
In the case of the Jaynes-Cummings model we could solve for the time evolution exactly. This is rarely the case and we must resort to perturbation theory in most cases.
Consider to start the definition of thw work:
\begin{equation}
    W=\tr(\rho_0 H_0)-\tr(U_I \rho_0 U^\dagger_I H_0)
\end{equation}
where we consider a Dyson expansion. In this case, the solution is given by the Dyson time ordering
\begin{equation}
    U_I(t)=\mathcal T \sum_{k=0}^\infty \frac{(-i)^k}{k!} (\int_0^t dt^\prime V_I(t^\prime))^k.
\end{equation}
We are interested in the case in which we need to resort to perturbation theory to evaluate the unitary operator above. Up to the second order, we have
\begin{eqnarray}
    U_2&=& G^\dagger \Big(\mathbb I -i \int^t_{t_0}  V_0(t^\prime)  dt^\prime%\nonumber \\
    %&\ &\ \ \ 
    -\frac{1}{2}   \int^t_{t_0}\int^t_{t_0}:  V_0(t^\prime) V_0(t^{\prime \prime}):  dt^\prime dt^{\prime \prime}\Big)G+ O(t^3),\nonumber \\
    U^\dagger_2& =& G^\dagger(\mathbb I +i \int^t_{t_0}  V_0^\dagger(t^\prime)  dt^\prime %\nonumber \\
%    &\ &\ \ \ 
    -\frac{1}{2}   \int^t_{t_0}\int^t_{t_0}(:   V_0(t^\prime)  V_0(t^{\prime \prime}) :)^\dagger  dt^\prime dt^{\prime \prime})G+O(t^3).   %\nonumber \\ 
\end{eqnarray}
In what follows, we can assume that $V_0^\dagger=V_0$. Given the expressions above, we have now to evaluate the average of
\begin{equation}
    W=\tr(\rho_0 H_0)-\tr(G^\dagger U_2 G \rho_0 G ^\dagger U^\dagger_2 G H_0)
\end{equation}
using the average of the unitary matrix $G$:
\begin{equation}
    \langle (G^\dagger \otimes G^\dagger)(U_2 \otimes U_2^\dagger) (G \otimes G)\rangle_G=\lambda_+ \Pi_++\lambda_- \Pi_-,
\end{equation}
with 
\begin{eqnarray}
    \lambda_+&=&\frac{\tr((U_2 \otimes U_2^\dagger)\Pi_+)}{\tr(\Pi_+)} %\nonumber \\    &=&
    =\frac{2}{n(n+1)} \frac{\tr(U_2)\tr(U_2^\dagger)+\tr(U_2 U_2^\dagger)}{2} \nonumber \\
    \lambda_-&=&\frac{\tr((U_2 \otimes U_2^\dagger)\Pi_-)}{\tr(\Pi_-)} %\nonumber \\
   % &=&
    =\frac{2}{n(n-1)} \frac{\tr(U_2)\tr(U_2^\dagger)-\tr(U_2 U_2^\dagger)}{2} \nonumber.
\end{eqnarray}
Note that $\text{Tr}(U_2 U_2^\dagger)=\text{Tr}(U_2^\dagger U_2)=n+O(t^3)$.  We can use at this point the eqns. (\ref{eq:eigs}) again.
After a rapid calculation we see that (up to corrections of order $t^3$), we have
\begin{equation}
    \lambda_{\pm}=\frac{n^2\pm n -n \tr(A^2)-\tr(A)^2}{n^2\pm n}
\end{equation}
and thus
\begin{eqnarray}
    \frac{\lambda_++\lambda_-}{2}&=&\frac{n^2-1+\text{Tr}(A)^2-n \text{Tr}(A^2) }{n^2-1} \nonumber \\
    \frac{\lambda_+-\lambda_-}{2}&=&-\frac{\text{Tr}(A)^2-n  \text{Tr}(A^2)}{n (n^2-n)}
\end{eqnarray}
where $A=\int^t_{t_0}  V_0(t^\prime)  dt^\prime$, where we used the fact that inside the traces one has $\tr(\int^t_{t_0}\int^t_{t_0}(:   V_0(t^\prime)  V_0(t^{\prime \prime}) :)^\dagger  dt^\prime dt^{\prime \prime})=\tr(\int^t_{t_0}\int^t_{t_0}:   V_0(t^\prime)  V_0(t^{\prime \prime}) :  dt^\prime dt^{\prime \prime} )$.
We can now write
\begin{eqnarray}
    \langle W\rangle_G&=&\tr(\rho_0 H_0)(1-\frac{\lambda_++\lambda_-}{2}
)- \frac{\lambda_+-\lambda_-}{2} \tr(H_0)\tr(\rho_0) \nonumber \\
&=&\tr(\rho_0 H_0)(1-\frac{ \left(\tr(A)^2-n\tr(A^2) +n^2-1\right)}{n^2-1} 
)% \nonumber \\
%&+& 
+\frac{ \tr(A)^2 - n\tr(A^2) }{n^2 - 1}  \frac{\tr(H_0)}{n}.\nonumber \\
&=& \frac{ \tr(A)^2 - n\tr(A^2) }{n^2 - 1}  \frac{\tr(H_0)}{n} %\nonumber \\
%&-&
-\tr(\rho_0 H_0)\frac{ \left(\tr(A)^2-n\tr(A^2) \right)}{n^2-1} \nonumber \\
&=& \frac{ \tr(A)^2 - n\tr(A^2) }{n^2 - 1}  \left(\frac{\tr(H_0)}{n}- \tr(\rho_0 H_0)\right) \nonumber \\
&=& \langle \Delta A^2 \rangle_G \langle E\rangle_{G} 
\end{eqnarray}
As it could be seen from the beginning, we see again explicitly that the average work is the product of two terms, the first is adimensional and due to the perturbation,
\begin{equation}
    \langle \Delta A^2 \rangle_G=\frac{ \tr(A)^2 - n\tr(A^2) }{n^2 - 1}
\end{equation}
and the second term has the dimensions of energy, and due to the density matrix only:
\begin{equation}
    \langle E\rangle_{G}=\left(\frac{\tr(H_0)}{n}- \tr(\rho_0 H_0)\right).
\end{equation}
This shows that no work can extracted if the density matrix is the one of a completely mixed state.
\subsubsection{Example: Harmonic perturbations}
Let us now consider the example of a $n$-level system. At time $t=0$, the system is described by the eigenvalue equation
\begin{equation}
    \hat H_0 \psi_m= E_m \psi_m,
\end{equation}
and thus the wavefunction as a function of time can be written as
\begin{equation}
\psi(t)=\sum_m c_m e^{- i  E_m t} \psi_m.    
\end{equation}
We consider now a harmonic perturbation of the form:
\begin{equation}
    \hat H_1(t)= \hat V e^{i \omega t}+\hat V^\dagger e^{-i \omega t}
\end{equation}
where $\hat V$ is a generic operator and $\hat V^\dagger$ its hermitean conjugate.

Then, according to the formulae we have derived, the average work if we consider random rotations with respect to $G$ of $\hat H_1(t)\rightarrow G^\dagger \hat H_1(t) G$, depends on 
\begin{eqnarray}
A&=& \int_{t_0}^t dt^\prime \left(\hat V e^{i \omega t^\prime}+\hat V^\dagger e^{-i \omega t^\prime}\right) \nonumber \\
&i&\hat V\frac{ \left(e^{i t_0 \omega }-e^{ i t \omega }\right)}{\omega }-i\hat V^\dagger \frac{ \left(e^{-i t_0 \omega }-e^{-i t \omega }\right)}{\omega } \nonumber \\
A^2&=&- \hat V^2 \frac{ \left(e^{i t_0 \omega }-e^{i t \omega }\right)^2}{\omega^2 }-(\hat V^\dagger)^2 \frac{ \left(e^{-i t_0 \omega }-e^{-i t \omega }\right)^2}{\omega^2 }\nonumber \\
&+&(\hat V \hat V^\dagger +\hat V^\dagger \hat V)\frac{ \left(e^{-i t_0 \omega }-e^{-i t \omega }\right)\left(e^{i t_0 \omega }-e^{i t \omega }\right) }{\omega^2 }
\end{eqnarray}

We now use:
\begin{eqnarray}
    \left(e^{-i t_0 \omega }-e^{-i t \omega }\right)&=& -2i e^{-i \frac{t+t_0}{2} \omega} \sin(\frac{t-t_0}{2} \omega ) \nonumber \\
    \left(e^{i t_0 \omega }-e^{i t \omega }\right)&=&2i e^{i \frac{t+t_0}{2} \omega} \sin(\frac{t-t_0}{2} \omega ) \nonumber
\end{eqnarray}
and thus, if we define $f(t,\omega)=2\frac{\sin(\frac{t-t_0}{2} \omega ) }{\omega}$, we have
\begin{eqnarray}
A&=&  -f(t,\omega) \left(\hat V e^{i \frac{t+t_0}{2} \omega}+\hat V^\dagger e^{-i \frac{t+t_0}{2} \omega }\right) 
 \nonumber \\
A^2&=& \left(\hat V^2 e^{i (t+t_0) \omega }+(\hat V^\dagger)^2 e^{-i (t+t_0) \omega } -(\{V,V^\dagger\}) \right) f^2(t,\omega)\nonumber 
\end{eqnarray}
where $\{V,V^\dagger\}=V V^\dagger+V^\dagger V$. 
At this point we are ready to perform the traces.
First, we have that
\begin{equation}
    \tr(A)=- \left(\tr(\hat V) e^{i \frac{t+t_0}{2} \omega}+\tr(\hat V^\dagger) e^{-i \frac{t+t_0}{2} \omega }\right) f(t,\omega).
\end{equation}
Let $\lambda_k$ be the complex eigenvalues of $\hat V$ and $\sigma_k$ the singular values. Then, we have
\ba
    \tr(A)&=&- \left(\tr(\hat V) e^{i \frac{t+t_0}{2} \omega}+\tr(\hat V^\dagger) e^{-i \frac{t+t_0}{2} \omega }\right) f(t,\omega) \nonumber \\
    &=&-2 \sum_k \text{Re}\left(\lambda_k e^{i \frac{t+t_0}{2} \omega}\right)f(t,\omega).
\ea
meanwhile
\begin{equation}
    \tr(A^2)=-2\left( \sum_k\text{Re}\left(\lambda_k^2 e^{i \omega (t_0+t)}\right) -\sum_k \sigma_k \right) f^2(t,\omega)
\end{equation}
And thus the $A$ dependent part of the average work is given by
\begin{eqnarray}
    \langle \Delta A^2\rangle_G &=&\frac{2 f^2(t,\omega)}{(n^2-1)} \Big(\sum_{k,k^\prime} \text{Re}\left(\lambda_k e^{i \frac{t+t_0}{2} \omega}\right)\text{Re}\left(\lambda_{k^\prime} e^{i \frac{t+t_0}{2} \omega}\right) \nonumber \\
    &-&2n \sum_k\text{Re}\left(\lambda_k^2 e^{i \omega (t_0+t)}\right)+2n \sum_k \sigma_k   \Big)
    \label{eq:avw}
\end{eqnarray}
which is the expression for the performed work due to a harmonic perturbation.
{What we see is that the overall work is proportional to product of two functions, one is the square of function $f(t,\omega)=2 \frac{\sin(\frac{t-t_0}{2} \omega)}{\omega}$ and a factor which depends on the eigenvalues of the operator $\hat V$. The function $f$ is periodic with period $\frac{2\pi}{\omega}$ and has a maximum for $t_k=(4   k+1)\frac{\pi}{\omega }+t_0$. If $\hat V$ is self-adjoint, $\sigma_k=\lambda_k^2$, and we have in the parenthesis the function
\begin{equation}
 \sum_{k,k^\prime}\lambda_k \lambda_{k^\prime} \cos^2(\frac{t+t_0}{2}\omega)+2n\sum_k \lambda_k^2\left(1-\cos^2(\frac{t+t_0}{2}\omega)\right)   
\end{equation}
which can be rewritten as
\begin{equation}
    2 n \sum_k \lambda_k^2+\cos^2(\frac{t+t_0}{2} \omega) \sum_{k\neq k^\prime }\lambda_k \lambda_{k^\prime}
\end{equation}
If we introduce the constants $a_0,b_0,c_0$, the work is thus a function of the form :
\begin{equation}
    a_0 \sin^2(x-x_0)\left(c_0+d_0 \cos^2(x+x_0)\right)
\end{equation}
which is periodic. For $t\gg t_0$, the function above has two minima if $c_0<d_0$ and only one for $c_0>d_0$. However, it is not hard to see that $c_0>d_0$ is always true if 
\begin{equation}
    \text{Tr}(A^2)- \frac{\text{Tr}(A)^2}{2n+1}\geq 0
\end{equation}
is always true $\forall A$. However the identity above follows immediately from the fact that
\begin{equation}
    \text{Tr}\left(a A+b I \right)^2\geq 0
\end{equation}
is true for arbitrary $a,b\in \mathbb{R}$, and it follows from the choice $a=n$, $b=c_{\pm} \text{Tr}(A)$
with 
\begin{equation}
    c_{\pm}=-1\pm \sqrt{1-\frac{n}{2n+1}}.
\end{equation}
Thus, the work performed by a (random) harmonic perturbation of the form $ 2\hat V  \cos(\omega t)$ has always a single maximum
at $t_k=(2k+1) \frac{\pi}{\omega}$ on average. This can be interpreted as the fact that there are specific moments at which we stop our process to have performed the maximum amount of work on the battery. 
}

%\begin{figure}
 %   \centering
 %   \includegraphics[scale=0.6]{ptheory.pdf}
 %   \caption{Average work from eqn. (\ref{eq:avw}) for $n=10,50,100$ and $\omega=0.05$/}
 %   \label{fig:avw}
%\end{figure}

%%%%%%%%%%%%%%%%%%%%%%%%%%%%%%%%%%%

\subsection{Random spacing for CUE ensemble}\label{appendix8e}
Consider the following problem.
Given the function
\begin{equation}
    Q=\sum_{i=1}^n \sum_{j=i+1}^n \cos(\theta_i-\theta_j)
\end{equation}
with $\lambda_j=e^{i\theta_j}$, we ask what is the approximate value of $Q$ for a random matrix in the Circulant Unitary Ensemble (CUE). First, we note that we can write
\begin{eqnarray}
    Q&=&\sum_{k=1}^n \sum_{j=k+1}^n \cos(-i \log\lambda_k-\log \lambda_j) \nonumber \\
    &=&\sum_{k=1}^n \sum_{j=k+1}^n \cosh(\log (\frac{\lambda_k}{\lambda_j})) \nonumber \\
    &=&\sum_{k=1}^n \sum_{j=k+1}^n \cosh(\log (\frac{\lambda_k}{\lambda_j}))=\frac{1}{2} \sum_{k=1}^n \sum_{j=k+1}^n \left(\frac{\lambda_k}{\lambda_j}+\frac{\lambda_j}{\lambda_k} \right)\nonumber  
\end{eqnarray}
Let us define $r_k=\frac{\lambda_{k+1}}{\lambda_k}$. We then see that we can write
\begin{equation}
    \frac{\lambda_{k+t}}{\lambda_{k}}=\prod_{j=0}^{t-1} r_{k+j}
\end{equation}
and thus
\begin{eqnarray}
    Q&=&\frac{1}{2} \sum_{k=1}^n \sum_{j=k+1}^n \left( \prod_{i=j+1}^n r_i +\prod_{i=j+1}^n r_i^{-1}\right)
\end{eqnarray}
the average of $Q$, evaluated numerically, is provided in Fig. \ref{fig:cue}. We see that for large values of $n$ the peak of the distribution moves towards zero.

%\begin{figure}
%    \centering
%    \includegraphics[scale=0.35]{GUE.png}
%    \caption{Average of $Q$ over 1000 samples for random matrices in the Gaussian Unitary Ensembles of dimensions $n=100,500,1000$. The peak of the distribution converges to zero for larger values of $n$.}
%    \label{fig:gue}
%\end{figure}

%%%%%%%%%%%%%%%%%%%%%%%%%%%%%%%%%%%
% ADIABATIC BATTERIES
%%%%%%%%%%%%%%%%%%%%%%%%%%%%%%%%%%%
\subsection{Adiabatic Quantum Batteries}\label{appendix8f}
Here we give the details for the calculation of work fluctuations $\Delta W^2_{ad}$ for the adiabatic batteries. We first recall the calculation of the average.
Let us start from the following protocol.
The Hamiltonian, for $\alpha=0,1$, is written for an adiabatic transformation as 
\begin{equation}
    H_{\alpha}=\sum_{i=1}^R \epsilon^i_\alpha \Pi_\alpha^i.
\end{equation}
Consider $\epsilon^i(t):[0,1]\rightarrow \mathbb R$, with $\epsilon^i(0)=\epsilon^i_0$, $\epsilon^i(1)=\epsilon^i_1$. 
It can be shown that the evolution of the projector operators can be written as
\begin{equation}
\Pi_\alpha^i(t)=U_t \Pi_\alpha^i(0) U_t^\dagger.
\end{equation}
Thus, the time evolution of the Hamiltonian for an adiabatic system can be written as
\begin{equation}
    H(t)=\sum_{i=1}^R \epsilon^i(t) U_t \Pi_0^i U^\dagger _t,
\end{equation}
where the 
while the density matrix as $\rho(t)=\sum_i p_i U_t \Pi_0^i U^\dagger _t$.
It is important that the vector $d_i^\alpha\equiv(\text{Tr}(\Pi^j_\alpha)$ does not change with time, and thus can simply call $d_i$ these quantities, meanwhile $n$ is the dimension of the Hilbert space.

Because these relationships are in a way independent from the intermediate states, we simply write these expressions for $t=0$ and $t=1$ without loss of generality.
The work as 
\begin{eqnarray}
    W
    &=&\text{Tr}(\rho_0 H_0)-\text{Tr}(\rho_1 H_0) \nonumber \\
    &=& \sum_{i=1}^R\text{Tr}(p_i (\Pi_0^i-\Pi_1^i) H_0) \nonumber \\
    &=& \sum_{i,j=1}^Rp_i \epsilon^j_0\text{Tr}( (\Pi_0^i-\Pi_1^i) \Pi_0^j)
\end{eqnarray}
We now have $\Pi_\alpha ^i \Pi_\beta ^j= \delta^{ij}$ if $\alpha=\beta$, but otherwise they are not necessarily orthogonal. Let us write the work as
\begin{eqnarray}
    W&=&\sum_{i,j} p_i \epsilon^j_0 \left( \text{Tr}(\Pi^i_0\Pi^j_0)-\text{Tr}(\Pi^i_0\Pi^j_1) \right) \nonumber \\
    &=& \sum_{i,j} p_i \epsilon^j_0 \left( d_i \delta_{ij}-\text{Tr}(\Pi^i_0 G \Pi^j_0 G^\dagger) \right)
\end{eqnarray}
We can now perform the average over the unitary transformation $U$. We obtain
\begin{eqnarray}
    \overline W&=&\sum_{i,j} p_i \epsilon^j_0 \left( d_i \delta_{ij}-\text{Tr}(\Pi^i_0  \frac{d_j \mathbb I}{n}) \right) \nonumber\\
    &=&\sum_{i,j} p_i \epsilon^j_0 \left( d_i \delta_{ij}-  \frac{ d_i d_j }{n} \right).
\end{eqnarray}
Since we will need it for the calculation of the fluctuations, we note that
\begin{eqnarray}
    \overline{W}^2&=&\sum_{i,j,k,l} p_i \epsilon^j_0 p_k \epsilon^l_0\big(d_i d_k \delta_{ij}\delta_{kl}+ \frac{d_i d_j d_k d_l}{n^2} %\nonumber \\
  % &\ &\ \ \ \ \ \ \ \  
   -\frac{d_i d_j d_k \delta_{kl}+d_l d_k d_i \delta_{ij}}{n}\big).
\end{eqnarray}
Let us now calculate the fluctuations. The square of the work reads
\begin{eqnarray}
    W^2&=&\sum_{i,j,k,l} p_i \epsilon^j_0 p_k \epsilon^l_0 \left( d_i \delta_{ij}-\tr(\Pi^i_0 G \Pi^j_0 G^\dagger) \right) %\nonumber \\
%    &\ &\ \ \ \ \ \ \ \cdot 
    \left( d_k \delta_{kl}-\tr(\Pi^k_0 G \Pi^l_0 G^\dagger) \right) \nonumber \\
    &=&\sum_{i,j,k,l} p_i \epsilon^j_0 p_k \epsilon^l_0 \Big(d_i d_k \delta_{ij} \delta_{kl}- d_i \delta_{ij} \tr(\Pi^k_0 G \Pi^l_0 G^\dagger) \nonumber \\
    &\ &\ \ \ \ \ -d_k \delta_{kl} \tr(\Pi^i_0 G \Pi^j_0 G^\dagger)%\nonumber \\
%    &\ &\ \ \ \ \ 
    +\tr(\Pi^k_0 G \Pi^l_0 G^\dagger) \tr(\Pi^i_0 G \Pi^j_0 G^\dagger)\Big) \nonumber \\
    &=&\sum_{i,j,k,l} p_i \epsilon^j_0 p_k \epsilon^l_0 \Big(d_i d_k \delta_{ij} \delta_{kl}- d_i \delta_{ij} \tr(\Pi^k_0 G \Pi^l_0 G^\dagger) \nonumber \\
    &\ &\ \ \ \ \ -d_k \delta_{kl} \tr(\Pi^i_0 G \Pi^j_0 G^\dagger) \nonumber \\
    &\ &\ \ \ \ \ +\tr\Big((\Pi^k_0 \otimes \Pi^i_0) (G \otimes G)( \Pi^l_0 \otimes \Pi^j_0)(G^\dagger \otimes G^\dagger)\Big)\Big) \nonumber 
\end{eqnarray}
We can now perform the averages. We obtain
\begin{eqnarray}
\langle {W^2}\rangle_{ad} &=&\sum_{i,j,k,l} p_i \epsilon^j_0 p_k \epsilon^l_0 \Big(d_i d_k \delta_{ij} \delta_{kl}-(d_i \delta_{ij} \frac{d_k d_l}{n}+ d_k \delta_{kl} \frac{d_id_j}{n} )%\nonumber \\
    %&\ &\ \ \ \ \ 
    +\tr\big((\Pi^k_0 \otimes \Pi^i_0) (\lambda_+ \Pi_++\lambda_i \Pi_-\big)\Big)\nonumber \\
    &=&\bar W^2+\sum_{i,j,k,l} p_i \epsilon^j_0 p_k \epsilon^l_0 \Big( \tr\big((\Pi^k_0 \otimes \Pi^i_0) (\lambda_+ \Pi_++\lambda_i \Pi_-\big)-\frac{d_i d_j d_k d_l}{n^2} \Big)
\end{eqnarray}
where 
\be
\lambda_{\pm}=\frac{\tr( \Pi^l_0 \otimes \Pi_0^j) \Pi_{\pm})}{\tr{\Pi_{\pm}}}=\frac{d_l d_j\pm d_l \delta_{lj}}{ n(n\pm1)}
\ee
{}{
Let us focus on:
\begin{eqnarray}
    \text{Tr}\big((\Pi^k_0 \otimes \Pi^i_0) (\lambda_+ \Pi_++\lambda_i \Pi_-\big)\big)&=&\text{Tr}\big((\Pi^k_0 \otimes \Pi^i_0) %\nonumber \\
%    &\ & \ \  \cdot 
    ( \frac{\lambda_++\lambda_-}{2} \mathbb{I}+\frac{\lambda_+-\lambda_-}{2} \mathbb{T})\big) \nonumber \\
    &=& \frac{\lambda_++\lambda_-}{2} d_k d_i+\frac{\lambda_+-\lambda_-}{2} d_k \delta_{ki}\nonumber \\
\end{eqnarray}
We note that
\begin{eqnarray}
     \frac{\lambda_++\lambda_-}{2}&=&\frac{1}{2}\left(\frac{d_l d_j+ d_l \delta_{lj}}{ n(n+1)}+\frac{d_l d_j- d_l \delta_{lj}}{ n(n-1)} \right)=\frac{d_l \left(d\ d_j-\delta _{lj}\right)}{n \left(n^2-1\right)} \nonumber \\
\frac{\lambda_+-\lambda_-}{2}&=&\frac{1}{2}\frac{d_j d_l+d_l \delta _{lj}}{n (n+1)}-\frac{d_j d_l-d_l \delta
   _{lj}}{(n-1) n}=\frac{d_l \left(d \delta _{lj}-d_j\right)}{n \left(n^2-1\right)} \nonumber \\
\end{eqnarray}
from which we obtain:
\begin{eqnarray}
    \text{Tr}\big((\Pi^k_0 \otimes \Pi^i_0) (\lambda_+ \Pi_++\lambda_i \Pi_-\big)\big)&=&d_k d_l \Big(\frac{d_i \left(d\ d_j-\delta _{lj}\right)}{n(n^2-1)}%\nonumber \\
%    &\ &
    + \frac{\delta_{ki}
   \left(d\ \delta_{lj}-d_j\right)}{n\ (n^2-1)}\Big) \nonumber \\
   &=&\frac{d_i d_j d_k d_l}{n^2-1}   -\frac{d_i d_k d_l \delta _{lj}}{n
  \left(n^2-1\right)}\nonumber %\\
 % &-&
  -\frac{d_j d_k d_l \delta _{ki}}{n
   \left(n^2-1\right)}  +\frac{d_k d_l \delta _{ki} \delta _{lj}}{n^2-1}\nonumber
\end{eqnarray}
We use the result on $\bar W^2$, and thus
\begin{eqnarray}
    \overline{W^2}-\overline{W}^2&=&\sum_{i,j,k,l} p_i \epsilon^j_0 p_k \epsilon^l_0 \Big( \frac{d_i d_j d_k d_l}{n^2-1}   -\frac{d_i d_k d_l \delta _{lj}}{n
  \left(n^2-1\right)}%\nonumber %\\
 % &-&
  -\frac{d_j d_k d_l \delta _{ki}}{n
   \left(n^2-1\right)}  +\frac{d_k d_l \delta _{ki} \delta _{lj}}{n^2-1}-\frac{d_i d_j d_k d_l}{n^2} \Big)
   \end{eqnarray}
For the dimension of the Hilbert space $n\gg 1$, the terms of order $1/n^3$ go to zero faster than $1/n^2$, and we obtain
\begin{equation}
\overline{W^2}-\overline{W}^2\underbrace{=}_{n\gg1 }\frac{1}{n^2} \sum_{i,j,k,l} p_i \epsilon^j_0 p_k \epsilon^l_0 (d_k d_l \delta _{ki} \delta _{lj})=\frac{\text{Tr}(H_0 \rho_0)^2}{n^2} =\frac{ E_0^2}{n^2}
\end{equation}
which exhibits concentration.

Let us now look at bounds on the adiabatic work compared to the mean work for arbitrary random evolutions. We consider 
\begin{eqnarray}
    \langle W\rangle_{ad}=E_0-\sum_{ij} \frac{p_i \epsilon_0^j}{n} d_i d_j \\
    \langle W\rangle=E_0-\frac{\tr H_0}{n}=E_0-\sum_{ij} \frac{p_i \epsilon_0^j}{n}.
\end{eqnarray}
It is easy to see that
\begin{equation}
    \langle W\rangle_{ad}-\langle W\rangle=\text{Tr}(AB)
\end{equation}
where $A_{ij}=d_i d_j-1$ and $B=\frac{p_i \epsilon_j}{n}$. We now know that for $A$ nonnegative and $B$ arbitrary, we have
\begin{equation}
    \text{Tr}(AB)\leq \sigma_{max}(B)\text{Tr}(A)=\sigma_{max}(B) (\sum_i d_i^2-n)
\end{equation}
where $\sigma_{max}(B)$ is the spectral norm of the matrix $B$ \cite{tb}. The matrix $B=\frac{p_i \epsilon_j}{n}$ has only two eigenvalues since it is rank one, which are $0$ and $\frac{1}{n} \sum_{i} p_i  \epsilon_0^i$. Thus the spectral norm is $\sigma_{\max}(B)=\text{max}(0,\frac{1}{n} \sum_{i} p_i  \epsilon_0^i)$. We thus find that the maximum gain that one can has from degeneracy is 
\begin{equation}
    \langle W\rangle_{ad}-\langle W\rangle\leq \text{Tr}(\rho H_0)\frac{\sum_i d_i^2-n}{n},
\end{equation}
from which we obtain
\begin{equation}
    \langle W\rangle_{ad}\leq E_0 (1+c)-\frac{\tr(H_0)}{n}
\end{equation}
with $c=\frac{\sum_i d_i^2-n}{n}$.

\end{document}